\documentclass[aps,prd,twocolumn,floatfix,showpacs,superscriptaddress,nofootinbib]{revtex4-1}  
\usepackage{color}
\usepackage{amssymb}
\usepackage{amsmath}
\usepackage{mathrsfs,dsfont}
\usepackage{bm,hyperref}
\usepackage{graphicx}
\usepackage{subcaption}

\newcommand{\threej}[6]{{\left(\begin{array}{ccc} #1 & #2 & #3 \\ #4 & #5 & #6 \end{array}\right)}}
\newcommand{\sixj}[6]{{\left\{\begin{array}{ccc} #1 & #2 & #3 \\ #4 & #5 & #6 \end{array}\right\}}}
\newcommand{\tj}[6]{ \begin{pmatrix}
     #1 & #2 & #3 \\
     #4 & #5 & #6	 
   \end{pmatrix}} 
\newcommand{\sj}[6]{ \begin{Bmatrix}
     #1 & #2 & #3 \\
     #4 & #5 & #6	 
\end{Bmatrix}} 

\newcommand{\changetext}[1]{#1}

\newcommand{\sk}[1]{}

\newcommand{\coefalpha}{0.75}

\newcommand{\sovast}{Sov. Astron.}
\newcommand{\mnras}{Mon. Not. R. Astron. Soc.}
\newcommand{\aj}{Astron. J.}
\newcommand{\apjl}{Astrophys. J. Lett.}
\newcommand{\jcap}{J. Cosmo. Astropart. Phys.}
\newcommand{\aap}{Astron. Astrophys.}

\begin{document}
\widetext

\title{Detecting primordial gravitational waves with circular polarization of the 
redshifted 21 cm line: I. Formalism}
\author{Christopher M. Hirata}
\email{hirata.10@osu.edu}
\affiliation{Center for Cosmology and Astro Particle Physics (CCAPP),
The Ohio State University,  191 West  Woodruff Lane, Columbus, Ohio 43210, USA}
\author{Abhilash Mishra}
\email{abhilash@astro.caltech.edu}
\affiliation{Theoretical Astrophysics Including Relativity (TAPIR),\\
Caltech, M/C 350-17, Pasadena, California 91125, USA}
\author{Tejaswi Venumadhav}
\email{tejaswi@sns.ias.edu}
\affiliation{School of Natural Sciences, Institute for Advanced Study, 1 Einstein Drive, Princeton, New Jersey 08540, USA}

\begin{abstract}
We propose a new method to measure the tensor-to-scalar ratio $r$ using the circular polarization of the 21 cm radiation from the pre-reionization epoch. Our method relies on the splitting of the $F = 1$ hyperfine level of neutral hydrogen due to the quadrupole moment of the CMB. We show that unlike the Zeeman effect, where $M_{F}=\pm 1$ have opposite energy shifts, the CMB quadrupole shifts $M_{F}=\pm 1$ together relative to $M_{F}= 0$. This splitting leads to a small circular polarization of the emitted 21 cm radiation.  In this paper (Paper I in a series on this effect), we present calculations on the microphysics behind this effect, accounting for all processes that affect the hyperfine transition. We conclude with an analytic formula for the circular polarization from the Dark Ages as a function of pre-reionization parameters and the value of the remote quadrupole of the CMB. We also calculate the splitting of the $F = 1$ hyperfine level due to other anisotropic radiation sources and show that they are not dominant. In a companion paper (Paper II) we make forecasts for measuring the tensor-to-scalar ratio $r$ using future radio arrays. 
\end{abstract}

\maketitle

\section{Introduction}
One of the major programs in modern cosmology is the search for primordial gravitational waves from inflation. Inflation is the leading paradigm for the solution to the horizon and flatness problems, and for the origin of large-scale structure, which is ascribed to quantum perturbations in the early Universe \cite{1981PhRvD..23..347G, 1982PhLB..114..431L, 1982PhRvL..48.1220A}. In addition to scalar (or density) perturbations, inflation also predicts a spectrum of tensor (or gravitational wave) perturbations, whose amplitude is directly related to the Hubble rate during inflation. A confirmed detection of the tensor perturbations would be a major victory for inflation, and characterization of the background would open a new window into the earliest fraction of a second of cosmic history.

At present, the most advanced probe of the primordial tensor perturbations is the polarization of the cosmic microwave background (CMB) \cite{1985SvA....29..607P, 1993ApJ...417L..13C, 1993PhLB..319...96H, 1995ApJ...445..521N}: the quadrupole anisotropy in the CMB induced by the shear strain of the gravitational wave is transformed via Thomson scattering into a polarization signal. In the late 1990s, it was found that one type of polarization mode in the CMB -- the $B$-mode -- is not sourced at linear order by scalar perturbations and is thus a potentially clean probe of the tensors \cite{1997PhRvD..55.1830Z, 1998PhRvD..57..685K}. The $B$-mode polarization is expected to show two peaks due to the two visible epochs with free electrons available: a ``recombination'' peak at angular scales of a few degrees, and a ``reionization'' peak at angular scales of several tens of degrees. 

In 2014, the BICEP2 experiment reported a detection of degree-scale $B$-mode polarization consistent with a tensor-to-scalar ratio of $r\approx 0.2$ \cite{2014PhRvL.112x1101A}. However, joint analyses of the BICEP2 and Planck data revealed that the observed $B$-mode polarization could be attributed to Galactic dust~\cite{2014JCAP...08..039F, 2014JCAP...10..035M, 2015PhRvL.114j1301B, 2016A&A...586A.133P}. The present upper bound on $r$ from combinations of multi-frequency CMB polarization data with more model-dependent constraints is $r<0.07$ (95\%\ CL) \cite{2016PhRvL.116c1302B}.

While the near term agenda for primordial gravitational wave studies focuses on the $B$-mode polarization produced by linear theory, there remains a great deal of interest in other potential methods to probe primordial gravitational waves, both to confirm this interpretation of a detected polarization signal (as opposed to other exotica that can produce $B$-modes \cite{2008PhRvL.100m1302J}) and to measure its spectral properties such as the tensor spectral index $n_{\rm t}$. Several proposals are based on conventional CMB and large-scale structure observables. One is to use higher-order correlation functions of the galaxy and arcminute-scale CMB polarization fields \cite{2012PhRvD..85l3540A}. Another is to use the tidal alignment of galaxies observed in weak lensing surveys \cite{2012PhRvD..86h3513S, 2014PhRvD..89h3507S, 2014arXiv1406.4871C} (the detection of the weak lensing shear $B$-mode in next-generation galaxy surveys seems unpromising \cite{2003PhRvL..91b1301D}). Yet another idea is to use the frequency dependence of Rayleigh scattering immediately after hydrogen recombination to obtain an additional set of $B$-mode multipoles \cite{2013JCAP...08..053L}. In the longer term, the community has realized that other observables have the potential to either probe much smaller values of $r$ and/or extend the range of wave numbers $k$, if the daunting technical challenges can be addressed. At the high-$k$ end of the spectrum, direct detection of inflationary gravitational waves with a network of laser interferometers may be possible \cite{2001CQGra..18.3473C, 2001PhRvL..87v1103S, 2006CQGra..23.2435C}. At the cosmic scale, the ultimate probe of large scale structure (in terms of the number of modes available) is provided by the redshifted 21 cm line of hydrogen \cite{2004PhRvL..92u1301L}. 
A very futuristic cosmic variance-limited 21 cm experiment sensitive to gravitational waves down to $r\sim 10^{-9}$ via gravitational lensing and intrinsic alignment effects on the local power spectrum \cite{2010PhRvL.105p1302M, 2012PhRvL.108u1301B}.

In a series of earlier papers~\cite{2017PhRvD..95h3010V,2017PhRvD..95h3011G} we considered the effect of primordial magnetic fields on the statistics of 21 cm radiation, which arises due to the Zeeman splitting of the $F=1$ hyperfine excited level of hydrogen. \changetext{(We use the conventional quantum number $F$ for the total angular momentum of the atom, including electron and nuclear spins.)} We showed that this splitting changes the angular distribution of emitted radiation from atoms that were excited by an anisotropic radiation field, and hence leads to a characteristic correction to the observed brightness temperature at $21(1+z)$ cm at second order in the optical depth $\tau$. During the course of our investigations, we learned that the CMB anisotropy also leads to a splitting of the $F=1$ level, but that the splitting has different symmetry properties (in the Zeeman effect the $M_F=\pm1$ levels have opposite energy shifts, whereas the CMB quadrupole shifts $M_F=\pm1$ together relative to $M_F=0$). This leads to a qualitatively different outcome: whereas the Zeeman effect on the 21 cm line leads to an anisotropic temperature power spectrum, the CMB anisotropy results in a small circular polarization in the 21 cm line. This circular polarization is in principle observable to us today since (unlike linear polarization) it is not scrambled by Faraday rotation during its propagation through the Milky Way (and possibly the intergalactic medium). The purpose of this paper (``Paper I'' in this series) is to report on the calculation of the microphysics  of this effect. In a companion paper (``Paper II'') we assess the detectability of primordial gravitational waves through this novel channel and discuss the foreground challenges. 

This paper is organized as follows: we outline the formalism used for our calculation in Sec.~\ref{sec:formalism}. \sk{The key calculation of this paper is outlined in Sec.~\ref{sec:magneticsplitting}, where we compute the relative change in hyperfine energy levels due to an anisotropic photon bath and provide a semi-classical explanation of the effect.} \changetext{We compute the relative change in the hyperfine energy levels due to an anisotropic photon bath in Sec.~\ref{sec:magneticsplitting}. The key calculation of this paper is contained in Sec.~\ref{sec:precession}, in which we compute the orientation of the hydrogen spins due to the precession associated with the energy splitting, and provide a semi-classical explanation of the effect.} We calculate the effect of the remote CMB quadrupole on 21 cm polarization in Sec.~\ref{sec:poleffect}, and we summarize our results in Sec.~\ref{sec:discussion}.

\changetext{\section{Background and formalism}
\label{sec:formalism}}

\changetext{The signal described in this paper has three major ingredients. First, the hydrogen atom spins are aligned by short-wavelength density perturbations due to the finite optical depth in the 21 cm line. Second, the spins precess in the background CMB quadrupole (which is nearly constant over scales much larger than the density perturbations). Finally, the decay of the spin-polarized upper state of the hydrogen atoms produces polarized 21 cm radiation. This section summarizes the formalism of Venumadhav et~al.~\cite{2017PhRvD..95h3010V} (hereafter V17) for describing these processes.}

\changetext{\subsection{Atomic spin polarization}}

The distribution of the hydrogen atoms among the 4 hyperfine states $|FM_F\rangle$ is described by a quantum mechanical density matrix, as described in \changetext{V17}. Averaged over timescales longer than $2\pi/\omega_{\rm hf} \sim 0.7$ ns, the correlation between the $|a\rangle = |00\rangle$ state and the three $F=1$ states becomes zero. However, we must fully describe the $3\times 3$ sub-block \changetext{$\rho_{M_FM'_F}$} corresponding to the degenerate $F=1$ states; for this purpose, we use the spherical components
\begin{eqnarray}
\mathscr{P}_{jm} &=& \sqrt{3(2j+1)} \sum_{m_1m_2} (-1)^{1-m_2} 
\nonumber \\ && \times \threej1j1{-m_2}m{m_1}\rho_{m_1m_2}, \label{eq:pjm}
\end{eqnarray}
defined for $j=0,1,2$ and $-j\le m\le j$. There are 9 independent real numbers here since $\mathscr P_{j,-m} = (-1)^m \mathscr P_{jm}^\ast$. The probability of being in the $F=1$ state is the trace of the $F=1$ sub-block, which is equal to $\mathscr P_{00}$, and the probability of being in the $|00\rangle$ state is given by $\rho_{aa} = 1-\mathscr P_{00}$. Note that the density matrix transforms like the expectation value of a spherical operator, not a state: for example, under a right-handed active rotation of the system by angle $\alpha$ around the $z$-axis, $\mathscr P_{jm}$ acquires a factor of $e^{im\alpha}$, not $e^{-im\alpha}$.

The orientation and alignment of the hydrogen atom spins is described by the $j=1$ and $j=2$ components of $\mathscr P_{jm}$, respectively. In particular, we note that the mean spin of a hydrogen atom is
\begin{equation}
\langle {\bm F} \rangle= \frac\hbar{\sqrt3} \left[
-2 (\Re \mathscr P_{11}\,\hat{\bm e}_x + \Im \mathscr P_{11}\,\hat{\bm e}_y)
+  \sqrt2 \mathscr P_{10}\,\hat{\bm e}_z \right],
\end{equation}
where the symbols $\Re$ and $\Im$ denote the real and imaginary parts, respectively. The \changetext{alignment} or quadrupole moment of the distribution of spins is
\begin{eqnarray}
&& \!\!\!\!\!\!\!\!
\langle F_{\langle\mu}F_{\nu\rangle}\rangle = \frac{\hbar^2}{\sqrt3}
\nonumber \\ && \!\! \times
\left(\begin{array}{ccc}
\!\!-\sqrt{\frac16}\mathscr P_{20} + \Re\mathscr P_{22}\!\!\!\!
 & \Im\mathscr P_{22} & -\Re\mathscr P_{21} \\
\Im\mathscr P_{22}
  & \!\!\!\!-\sqrt{\frac16}\mathscr P_{20} - \Re\mathscr P_{22} & - \Im\mathscr P_{21} \\
  -\Re\mathscr P_{21} & -\Im\mathscr P_{21} & \sqrt{\frac23}\mathscr P_{20}
\end{array}\right),
\nonumber \\ &&
\end{eqnarray}
where the $\langle\rangle$ brackets in the subscript denote the \changetext{traceless-symmetric} part.

\changetext{\subsection{Alignment by density perturbations}}

\changetext{In a homogeneous and isotropic Universe, the hydrogen atom spins in the pre-reionization gas would point in random directions on average, i.e.\ $\mathscr P_{jm}=0$ for $j\neq 0$. A density perturbation in the pre-reionization cosmic gas, however, results in a local velocity gradient, $\partial_iv_j$, that has a traceless-symmetric part. This means that the optical depth of the 21 cm line is less than the cosmic mean optical depth for photons traveling in the ``compressing'' direction, and is greater for photons traveling in the ``stretching'' direction. Thus, the 21 cm radiation field incident on the hydrogen atoms is anisotropic; due to the transverse nature of electromagnetic waves, there is an associated anisotropy in the local magnetic field that is responsible for exciting the atoms to the $F=1$ level. The hydrogen atoms thus develop a spin polarization, with $j=2$ symmetry ($\mathscr P_{2m}\neq 0$) since it is sourced by a $j=2$ perturbation to the velocity gradient.}

\changetext{The alignment so produced for a single Fourier mode has been calculated in Eq.~(96) of V17:
\begin{eqnarray}
\mathscr P_{2m}({\bm k}) &=& \frac1{20\sqrt2} \frac{T_\star}{T_\gamma}\left(1-\frac{T_\gamma}{T_s}\right)\frac{\tau}{1 + x_{\alpha,(2)} + x_{c,(2)}}
\nonumber \\
&&\times f \, \delta({\bm k}) \sqrt{\frac{4\pi}5}Y_{2m}({\hat{\bm k}}), \label{eq:PI:2m}
\end{eqnarray}
where ${\bm k}$ is the Fourier wave vector; $\delta$ is the matter density perturbation; $f$ is the rate of growth of structure, and is $\approx 1$ in the matter-dominated era; $Y_{2m}$ is a spherical harmonic; $\tau$ is the cosmic mean optical depth in the 21 cm line; $T_\star = 68$\ mK is the hydrogen hyperfine splitting in temperature units; $T_\gamma = 2.725(1+z)$\ K is the CMB temperature; $x_{\alpha,(2)}$ and $x_{c,(2)}$ are coefficients describing the rate of de-alignment of polarized hydrogen atoms (these are dimensionless because they are defined relative to the rate of stimulated emission via the CMB); and we have dropped the term involving primordial magnetic fields (assumed negligible in this paper). Note that the alignment is proportional to the 21 cm optical depth, and would vanish in the case of $T_s=T_\gamma$.
Furthermore, the alignment is in the direction of ${\bm k}$, as must be true for a linear scalar perturbation.}

\changetext{The net spin orientation of $j=1$ symmetry ($\mathscr P_{1m}$ or $\langle{\bm F}\rangle$) must be zero for linear scalar perturbations because it transforms as an axial vector. Even going beyond linear perturbation theory, 21 cm absorption and emission can {\em only} source $\mathscr P_{1m}$ if there is incident circularly polarized radiation (see Eqs.~34, 37, and B12b of Ref.~\cite{2017PhRvD..95h3010V}). Thus we conclude that if only the conventional mechanisms are included, then $\mathscr P_{1m}=0$.
}

\changetext{\subsection{Precession in an anisotropic background}}

\changetext{Hydrogen atoms that are aligned (in the $j=2$ or ``headless-vector'' sense) emit linearly but not circularly polarized radiation. As we will see in Sec.~\ref{ss:emcpol}, we need $\mathscr P_{1m}\neq 0$ to produce circular polarization.}

\changetext{If the hydrogen atoms are subjected to an anisotropic perturbation that lifts the degeneracy of the three $F=1$ states, then they will precess and $\rho_{M_FM'_F}$ will change. Even a very weak perturbation will suffice: an $F=1$ hydrogen atom has a lifetime of $t_{\rm d}\sim T_\star/(AT_\gamma)\sim 10$\ kyr (where $A$ is the Einstein coefficient), and so order-unity precession angles could be realized if the energy levels shift by $\sim\hbar/10$\ kyr. This was the key idea behind the search for ultra-weak magnetic fields with cosmological 21 cm radiation \cite{2017PhRvD..95h3010V}.}

\changetext{Quantum mechanically, precession of a spin-$F$ system is described by a Hermitian perturbation Hamiltonian $\Delta{\cal E}_{M_FM'_F}$ where $M_F,M'_F\in\{-F...+F\}$. The precession causes the density matrix evolves according to $\dot\rho|_{\rm prec} = i[\rho,\Delta{\cal E}]$. Like the density matrix, the perturbation Hamiltonian has spherical components that transform as $j=0, 1, ...2F$ (in our case: $j=0$, 1, and 2). The monopole ($j=0$) part of the perturbation Hamiltonian corresponds to an overall shift of the energy levels and causes no precession. The dipole ($j=1$) part of the perturbation Hamiltonian would be sourced by an external magnetic field (the Zeeman effect). It results in solid-body rotation of the atomic density matrix, but this does not convert an alignment $\mathscr P_{2m}$ into an orientation $\mathscr P_{1m}$.}

\changetext{The main effect of interest in this paper is that the CMB anisotropy can source the quadrupole ($j=2$) part of the perturbation Hamiltonian. External perturbations with $j\ge 2$ lead to more complicated evolution of the atomic density matrix since the precession is not a solid-body rotation: in particular, they can inter-convert alignment $\mathscr P_{2m}$ and orientation $\mathscr P_{1m}$. This means that in the presence of a CMB quadrupole, the density perturbations can align the hydrogen spins by radiative transfer effects, and then precession can lead to a net spin orientation $\langle{\bm F}\rangle \neq 0$. Section \ref{sec:precession} provides a detailed calculation of this effect.}

\changetext{\subsection{Emission of circularly polarized radiation}
\label{ss:emcpol}}

\changetext{When a hydrogen atom decays from $F=1$ to $F=0$, its angular momentum is transferred to the emitted 21 cm photon. Therefore, if an observer views the gas along the $z$-direction, the observer will see a net circular polarization proportional to $F_z$ or $\mathscr P_{10}$.}

\changetext{Following the formalism of Ref.~\cite{2017PhRvD..95h3010V}, we describe circular polarization in terms of the multipole moments of the photon phase space density:
\begin{eqnarray}
&& \frac{f_{++}({\bm r},\omega,\hat{\bm n})
- f_{--}({\bm r},\omega,\hat{\bm n})}2
\nonumber \\
&& ~~~~=
\sum_{j=0}^\infty \sum_{m=-j}^j \sqrt{\frac{4\pi}{2j+1}} f_{{\rm V}, jm}({\bm r},\omega) Y_{jm}^\ast(\hat{\bm n}), \label{eq:fvdef}
\end{eqnarray}
where $f_{++}$ and $f_{--}$ are the phase space densities for right- and left-circularly polarized radiation at angular frequency $\omega$, direction $\hat{\bm n}$, and position ${\bm r}$. Since there is a rapid change in the solution at the 21 cm line itself, the independent variable $\omega$ is usually replaced by the cumulative line profile ${\mathcal X} = \int \phi(\omega)\,d\omega$, where $\phi(\omega)$ is the line profile (e.g.\ a Gaussian for a thermally broadened line); we have ${\mathcal X}=0$ on the red side of the line and ${\mathcal X}=1$ on the blue side. The radiative transfer equation is solved with the boundary condition of pure CMB (i.e.\ no circular polarization) at ${\mathcal X}=1$, in much the same way as the original Sobolev line transfer problem \cite{1960mes..book.....S}. The properties of the emitted radiation are extracted at ${\mathcal X}=0$, and then transformed into observable quantities by assuming free-streaming (phase space density conserved along a geodesic).
}

\changetext{We will show in Sec.~\ref{ss:4b} that in the limit of $\tau\ll 1$, the circular polarization on the red side of the line is
\begin{equation}
f_{{\rm V},1m}({\mathcal X}=0) = \sqrt{\frac83}\,\frac{T_\gamma T_s}{T_\star^2} \tau \mathscr P_{1m}.
\end{equation}
All aspects of this equation except for the numerical pre-factor could be anticipated based on simple physical arguments.
It is proportional to the net spin $\langle{\bm F}\rangle$ of the hydrogen atoms and the optical depth $\tau$, and has a dipolar form (if the radiation is right circularly polarized as seen from one direction, it is left circularly polarized as seen from the opposite direction). The factor of $T_\gamma/T_\star$ is the stimulated emission factor. Finally, the factor of $T_s/T_\star$ is the Rayleigh-Jeans phase space density that one would expect for an optically thick line.\\} 

\section{Magnetic dipole splitting}
\label{sec:magneticsplitting}

\changetext{In this section, we will compute the perturbations to the sublevels of the $F=1$ level for neutral hydrogen atoms (in their ground electronic state) immersed in a possibly anisotropic CMB.}

There is a history of finite-temperature calculations of the shifts of energy levels of atoms. Much of this work focused on the second order electric dipole induced shift in the ground and excited states \cite{1952RSPSA.214..137A, 1971PhRvL..27..208W, 1972JPhA....5..417K, 1981PhRvA..23.2397F, 2008PhRvA..78d2504J}, and recently finite-temperature quantum electrodynamics has been applied \cite{2008PhRvA..78c2520E}. There has also been work on the effect of external fields (both static and dynamic) on the hyperfine splitting \cite{1965PhRv..137..702F, 1982PhRvA..25.1233I, 2006PhRvL..97d0801B, 2006PhRvL..97d0802A}. The energy shift caused by blackbody radiation has even been measured experimentally in alkali atom Rydberg states \cite{1984PhRvL..53..230H} and in the $^{133}$Cs hyperfine transition \cite{1997PhRvL..78..622B, 2014PhRvL.112e0801J, 2014PhRvL.112e0801J}. None of these calculations provides the quantity we need, which is the relative change in the sub level energies (e.g. ${\cal E}_{M_F=1}-{\cal E}_{M_F=0}$) due to an anisotropic external blackbody. That calculation is the subject of this section. While the discussion here is self-contained, it draws heavily on the methodology of the aforementioned references.

\subsection{Setup}

According to second-order perturbation theory, an interaction leads to a change in the Hamiltonian matrix element between two otherwise degenerate states:
\begin{equation}
\Delta {\cal E}_{ji} = \left\langle \langle j|H_{int}|i\rangle + \sum_{n,\Gamma}\frac{\langle j|H_{int}|n,\Gamma\rangle \langle n,\Gamma|H_{int}|i\rangle}{{\cal E}_j-{\cal E}_{n,\Gamma}} \right\rangle_{\rm rad},
\label{eq:aC1}
\end{equation}
where $H_{int}$ is the Hamiltonian of the interaction, $|n\rangle$ denotes an intermediate state of the hydrogen atom, $\Gamma$ denotes a state of the radiation field, and ${\cal E}_{n,\Gamma}$ is the energy of $|n\rangle$ plus the additional energy due to all photons present or absent in the intermediate state $|\Gamma\rangle$ relative to the initial radiation state. The expectation value is taken over statistical realizations of the radiation field.

Before we evaluate Eq.~(\ref{eq:aC1}), some simple comments are in order. We will consider here both the electric dipole and magnetic dipole interactions here, as it is not a priori obvious which dominates. In both cases, the interaction Hamiltonian can be written schematically in the form:
\begin{equation}
H_{int} = \sum_{mnA} C_{mnA} |m\rangle \langle n| a_A + {\rm h.c.},
\label{eq:aC2}
\end{equation}
where $a_A$ is a photon annihilation operator in mode $A$ (we have not chosen the planar or spherical basis yet) and $C_{mnA}$ is a set of coefficients. Since averaged over a wave period the annihilation operator has zero expectation value, the expectation value of the first-order term in Eq.~(\ref{eq:aC1}) vanishes and we are left with the second-order term. Conceptually, this is because the mean of the electric or magnetic field of the radiation vanishes at the position of the atom. The second-order term in Eq.~(\ref{eq:aC1}) with the schematic form of Eq.~(\ref{eq:aC2}) gives
\begin{eqnarray}
&& \!\!\!\!\!\!\!\!\!\!\!\!\!\!\!\!
\Delta {\cal E}_{ji} = \Delta {\cal E}_{ji}({\rm vac})
+ \sum_{n,AB} \langle a^\dagger_B a_A \rangle
\nonumber \\ && ~~\times
\left(
\frac{C_{jnA}C_{inB}^\ast }{{\cal E}_j-{\cal E}_{n}-\hbar\omega}
+ \frac{C_{njB}^\ast C_{niA}}{{\cal E}_j-{\cal E}_{n}+\hbar\omega}
\right),
\label{eq:aC3}
\end{eqnarray}
\changetext{where $\omega=\omega_A=\omega_B$ is the frequency of the background photon,} $\Delta{\cal E}_{ji}({\rm vac})$ is the energy shift in vacuum (the Lamb shift) and the radiation operators have been normal-ordered. The vacuum shift does not affect the degeneracy of the $M_F$ sub levels since it respects isotropy, and in any case is already included in the measured hyperfine frequency.


Since all of the states for which we are computing energy shifts have positive parity, the remaining two terms in Eq.~(\ref{eq:aC3}) can be broken down into two pieces: a magnetic dipole shift (where both $H_{int}$s are magnetic dipole operations) and an electric dipole shift (where both are electric dipoles); cross-terms must vanish by parity. The magnetic dipole interaction is weaker, but has a smaller energy denominator for illumination by CMB photons; thus we do not know without a calculation which contribution dominates. The magnetic dipole term is calculated in the main text, where it is shown that it dominates. The electric dipole term is calculated in Appendix \ref{sec:ed}.

\subsection{Computation of the magnetic dipole interaction}
\label{ss:md}

We first consider the case of a magnetic dipole, where the interaction of the atom with incident radiation is described by a magnetic-type transition dipole moment $\bm \mu$ and an interaction Hamiltonian $H_{int} = {\bm\mu} \cdot{\bm B}$. For incident radiation at frequency $\omega$,
\begin{eqnarray}
\!\!\!\!\!\!\!\!
\Delta {\cal E}_{ji}^{\rm m.d.} &=&
 \sum_{n\mu\nu} \langle :\!B^{{\rm rad}(-)}_\mu B^{{\rm rad}(+)}_\nu\!: \rangle
\nonumber \\ && \times
\left[
\frac{(\mu_\mu)_{ni}(\mu_\nu)_{jn} }{{\cal E}_j-{\cal E}_{n}-\hbar\omega}
+ \frac{(\mu_\nu)_{ni}(\mu_\mu)_{jn}}{{\cal E}_j-{\cal E}_{n}+\hbar\omega}
\right],
\label{eq:ESm.d.}
\end{eqnarray}
where $B^{\rm rad}_\mu$ is the $\mu$-component of the magnetic field associated with the radiation and the subscripts $\mu$ and $\nu$ are summed over the 3 coordinate axes. Pairs of Roman subscripts on the right-hand side are shorthand for matrix elements of the dipole moment between atomic states. The $^{(\pm)}$ superscripts denote the positive and negative-frequency components, and $::$ is a reminder of normal ordering (which is already satisfied in this case). 

The magnetic dipole operator from the $1s_{1/2}(F=1)$ level connects only to the $1s_{1/2}(F=0)$ and $1s_{1/2}(F=1)$ levels, with the electron magnetic moment dominating. As in \changetext{V17}, we use a lowercase roman ``$a$'' to denote the $1s_{1/2}(F=0)$ level, and the appropriate magnetic quantum number to denote sublevels of the $1s_{1/2}(F=1)$ level. The relevant matrix elements of the transition dipole moment are
\begin{align}
  (\mu_{\nu})_{a m} & = \frac{g_e \mu_B}{\hbar} \langle a \vert S_{e, \nu} \vert 1 m \rangle = \frac{g_e \mu_B}{2} e_{(m), \nu},
\end{align}
where $\mu_B$ is the Bohr magneton and $g_e \approx 2$ is the electron g-factor, and we used the notation for the helicity basis vectors
\begin{equation}
{\bm e}_{(0)} = {\bm e}_z ~~{\rm and}~~{\bm e}_{(\pm1)} = \mp\frac1{\sqrt2}({\bm e}_x \pm i{\bm e}_y).
\end{equation}
We find for the $F=1$ level that 
\begin{eqnarray}
\!\!\!\!\!\!\!\!
\Delta {\cal E}_{mm'}^{\rm m.d.} &=&
\frac{\mu_B^2}{\hbar}
\sum_{\mu\nu} \langle :\!B^{{\rm rad}(-)}_\mu B^{{\rm rad}(+)}_\nu\!: \rangle
\nonumber \\ && \times
\Bigl[
- \frac{\omega_{hf}}{\omega^2-\omega_{hf}^2} (e_{(m')\nu}e^\ast_{(m)\mu} + e_{(m')\mu}e^\ast_{(m)\nu})
\nonumber \\ &&
+ (\mu\nu\,{\rm antisym.})
\Bigr],
\label{eq:MDF1}
\end{eqnarray}
where ``$\mu\nu$ antisym.'' denotes terms antisymmetric in $\mu$ and $\nu$ that will not be needed.
Taylor-expanding to first order in ${\cal E}_j-{\cal E}_n$ (valid since CMB photon frequencies are much greater than $\omega_{hf}$), we can write
\begin{eqnarray}
\!\!\!\!\!\!\!\!
\Delta {\cal E}_{mm'}^{\rm m.d.} &=&
\frac{\mu_B^2}{\hbar\omega}
\sum_{\mu\nu} \langle :\!B^{{\rm rad}(-)}_\mu B^{{\rm rad}(+)}_\nu\!: \rangle
\nonumber \\ && \times
\Bigl[
- \frac{\omega_{hf}}{\omega} (e_{(m')\nu}e^\ast_{(m)\mu} + e_{(m')\mu}e^\ast_{(m)\nu})
\nonumber \\ &&
+ (\mu\nu\,{\rm antisym.})
\Bigr].
\end{eqnarray}
The magnetic field power spectrum $\langle :\!B^{{\rm rad}(-)}_\mu B^{{\rm rad}(+)}_\nu\!: \rangle$ is $\mu\nu$ symmetric if the CMB has only intensity and linear polarization (with negligible circular polarization). Therefore we drop the ``$\mu\nu$ antisym.'' term.
The angular anisotropy (including the quadrupole $\ell=2$, but not the dipole $\ell=1$ due to parity considerations) can contribute to the surviving term. The isotropic CMB background can contribute as well, but it shifts all three $M_F$ values by the same amount and so does not contribute to splitting.

We can now estimate the magnetic dipole energy shift caused by the CMB. For a blackbody at temperature $T_\gamma$, the radiative part of the magnetic field has a mean squared value $\langle :\!B^{{\rm rad}\,2}\!:\rangle = 4\pi a_{\rm rad}T_\gamma^4$, where $a_{\rm rad}$ is the radiation energy density constant. Using that half of the energy density is at positive frequency and half at negative, and that the mean square magnetic field is equally distributed on the 3 coordinate axes, we find
\begin{equation}
\langle :\!B^{{\rm rad}(-)}_\mu B^{{\rm rad}(+)}_\nu\!: \rangle = \frac23\pi a_{\rm rad}T_\gamma^4 \delta_{\mu\nu}.
\label{eq:Brms}
\end{equation}
The mean value of $\omega^{-2}$ over the spectrum, weighted by energy density, is $\langle \omega^{-2}\rangle = (5/2\pi^2) (k_BT_\gamma/\hbar)^{-2}$. Thus the CMB-induced energy shift is
\begin{eqnarray}
\Delta{\cal E}^{\rm m.d.}_{mm'} &=& -\frac{\mu_B^2 }{\hbar} \frac{5\omega_{hf}}{2\pi^2(k_BT_\gamma/\hbar)^2}
\frac{2\pi a_{\rm rad} T_\gamma^4}3
(2\delta_{mm'})
\nonumber \\
&=& -1.2\times 10^{-9}\,{\rm s}^{-1}\,\left( \frac{T_\gamma}{60\,\rm K} \right)^2 \hbar\delta_{mm'}.
\end{eqnarray}

The energy splitting between different values of $m$ arises from the quadrupole anisotropy in the CMB that causes the tensor $\langle :\!B^{{\rm rad}\,2}\!:\rangle$ to have a symmetric-traceless component. Let us consider a quadrupole anisotropy of the form $T({\mathbf n}) = T_\gamma[1 + a_{20}Y_{20}({\mathbf n})]$. Then by symmetry around the $z$-axis, the magnetic fields on the $x$, $y$, and $z$ axes are still uncorrelated, but the $z$ magnetic dipole sees a mean temperature of $T_\gamma[1-(20\pi)^{-1/2}a_{20}]$ and the $x$ and $y$ dipoles see a mean temperature of $T_\gamma[1+(80\pi)^{-1/2}a_{20}]$. This difference in temperatures leads to a difference in the energy given by
\begin{eqnarray}
\frac{\Delta{\cal E}^{\rm m.d.}_{11}-\Delta{\cal E}^{\rm m.d.}_{10}}{\hbar} \!\!&=&\!\! -\sqrt{\frac5{\pi^3}}
 \frac{\mu_B^2\omega_{hf} a_{\rm rad}T_\gamma^2}{k_B^2}\,a_{20}
~~~~\nonumber \\
\!\!&=&\!\! -4.4\times 10^{-10}\,{\rm s}^{-1}\,\left( \frac{T_\gamma}{60\,\rm K} \right)^2 a_{20}.~~~~~~
\label{eq:MDe}
\end{eqnarray}
For typical CMB quadrupole anisotropies of order $2\times 10^{-5}$, and temperatures of order 60 K ($z\sim 20$), the energy splitting and hence the precession rate is of order $10^{-14}$ s$^{-1}$.

\subsection{Generalization to arbitrary CMB anisotropy}

The calculation above is valid strictly only for the $a_{20}$ quadrupole moment of the CMB. However, it is easily generalized to other components. First, we recall that the tensor $\langle :\!B^{{\rm rad}(-)}_\mu B^{{\rm rad}(+)}_\nu\!:\rangle$ has spin-0, 1, and 2 parts, all of positive parity, and that symmetry requires that they can be contributed only by the CMB monopole (mean temperature), the circular polarization dipole ($\ell=1$), and the quadrupole anisotropy ($\ell=2$ $T$ or $E$-mode polarization) respectively. Neglecting the $E$-mode polarization in comparison to the much larger temperature quadrupole, we conclude that the energy shift matrix element due to the CMB anisotropies, $\Delta {\cal E}_{mm'}^{\rm m.d.}$, has the property
\begin{equation}
\Delta {\cal E}_{m,m'}^{\rm m.d.} \propto \sum_{m''} (-1)^{m} \threej121{-m}{m''}{m'} a_{2,m''}
\label{eq:magsym}
\end{equation}
on account of the Wigner-Eckart theorem.
Defining the combination of constants
\begin{equation}
K_{\rm mag} \equiv \sqrt{\frac{50}{3\pi^3}}
 \frac{\mu_B^2\omega_{hf} a_{\rm rad}T_{\gamma,0}^2}{k_B^2}
 = 1.65\times 10^{-12}\,{\rm s}^{-1},
\end{equation}
we may use Eq.~(\ref{eq:MDe}) to find the constant of proportionality in Eq.~(\ref{eq:magsym}):
\begin{eqnarray}
\Delta {\cal E}_{m,m'}^{\rm m.d.} &=&
\hbar K_{\rm mag}(1+z)^2 
\nonumber \\ && \times
\sum_{m''} (-1)^{m} \threej121{-m}{m''}{m'} a_{2,m''}.
\label{eq:magsym2}
\end{eqnarray}

\section{Effect on atomic density matrix}
\label{sec:precession}

\changetext{The main result of the previous section is Eq.~\eqref{eq:magsym2}, which is the perturbation to the sublevels of the $F=1$ level of neutral hydrogen atoms due to a quadrupolar CMB anisotropy. In this section, we will derive the effect of this perturbation on the density matrix of the hydrogen atoms.}

\changetext{\subsection{Computation of the change in density matrix}}

The \changetext{relevant} part of the atomic density matrix evolves in accordance with the energy shift $\Delta{\cal E}_{m,m'}$ just as it does with any other energy shift:
\begin{equation}
\dot\rho^{\rm shift}_{mm'} = i[\rho,\Delta{\cal E}]_{mm'}
=  i\rho_{mm_1}\Delta{\cal E}_{m_1m'}
-i\Delta{\cal E}_{mm_1}\rho_{m_1m'}. \label{eq:commutator}
\end{equation}
Using Eq.~(\ref{eq:magsym2}), we can determine the evolution of the spherical components of the density matrix,
\begin{widetext}
\begin{eqnarray}
\dot{\mathscr{P}}_{jm} &=& \sqrt{3(2j+1)} \sum_{m_1,m_2} (-1)^{1-m_2} \threej 1j1{-m_2}{m}{m_1} \dot\rho_{m_1,m_2}
\nonumber \\
&=& i \sqrt{3(2j+1)}\,K_{\rm mag}(1+z)^2
\sum_{m'',m_3,m_1,m_2} (-1)^{-m_2+m_1} \threej 1j1{-m_2}{m}{m_1}
\threej 121{-m_1}{m''}{m_3} a_{2,m''} \rho_{m_3,m_2} + {\rm h.c.}
\nonumber \\
&=& i \sqrt{2j+1}\,K_{\rm mag}(1+z)^2
\sum_{m''m_3j'm'm_1m_2} (-1)^{-1+m_1} \threej 1j1{-m_2}{m}{m_1}
\threej 121{-m_1}{m''}{m_3} a_{2,m''}
\nonumber \\ && \times
\sqrt{2j'+1} \threej 1{j'}1{-m_2}{m'}{m_3} \mathscr{P}_{j'm'}
 + {\rm h.c.}
\nonumber \\
&=& i \sqrt{2j+1}\,K_{\rm mag}(1+z)^2
\sum_{j'm'm''} (-1)^{1+j+j'+m'} \sqrt{2j'+1}
\sixj j2{j'}111 \threej j2{j'} {m}{m''}{-m'}
a_{2,m''} \mathscr{P}_{j'm'} + {\rm h.c.}
\nonumber \\
&=& i \sqrt{2j+1}\,K_{\rm mag}(1+z)^2
\sum_{j'm'm''} (-1)^{1+j+j'+m'} \sqrt{2j'+1}
\sixj j2{j'}111 \biggl[ \threej j2{j'} {m}{m''}{-m'}
a_{2,m''} \mathscr{P}_{j'm'}
\nonumber \\ &&
- (-1)^{m} \threej j2{j'} {-m}{m''}{-m'}
a_{2,-m''}^\ast \mathscr{P}_{j',-m'}^\ast
\biggr]
\nonumber \\
&=& 2i \sqrt{2j+1}\,K_{\rm mag}(1+z)^2
\sum_{j'm'm'',\,j+j'~\rm odd}
 \sqrt{2j'+1}
\sixj j2{j'}111
(-1)^{m'}\threej j2{j'} {m}{m''}{-m'}
a_{2,m''} \mathscr{P}_{j'm'}.
\label{eq:mag.Pjm.evol-temp1}
\end{eqnarray}
\end{widetext}
[Here ``h.c.'' denotes the addition of the same term but with the replacement $m\rightarrow -m$, a complex conjugate, and a factor of $(-1)^m$. In the last steps this term is evaluated, leading ultimately to a cancellation if $j+j'$ is even and a factor of 2 if $j+j'$ is odd. The unusual factor of $(-1)^{m'}$ rather than $(-1)^m$ is the result of the complex conjugation conventions: spherical tensor operators such as ${\mathscr P}_{jm}$ are defined to pick up a factor of $e^{im\alpha}$ under active right-handed rotation by $\alpha$ around the $+z$ axis, whereas the CMB multipole moments $a_{2m}$ are defined like coefficients of quantum states to have a factor of $e^{-im\alpha}$.]
Due to the $j+j'={\rm odd}$ rule and the triangle inequality, only the terms with $(j,j')=(1,2)$ or $(2,1)$ contribute, and the $6j$-symbols for these parameters evaluate to $-1/\sqrt{20}$. Thus we find that
\begin{subequations}
\label{eq:a-evolve}
\begin{align}
\dot{\mathscr{P}}_{00} &= 0, \\
\dot{\mathscr{P}}_{1m} &= -\sqrt{3}\,iK_{\rm mag}(1+z)^2
\sum_{m'm''}
\threej 122 {m}{m''}{-m'}
\nonumber \\ & \times
(-1)^{m'}a_{2,m''} \mathscr{P}_{2m'},~~~~{\rm and}
\label{eq:p1production} \\
\dot{\mathscr{P}}_{2m} &= -\sqrt{3}\,iK_{\rm mag}(1+z)^2
\sum_{m'm''}
\threej 221 {m}{m''}{-m'}
\nonumber \\ & \times
(-1)^{m'}a_{2,m''} \mathscr{P}_{1m'}.
\end{align}
\end{subequations}
[Since the precession does not move atoms into or out of the $F=1$ level, we already knew that $\dot{\mathscr{P}}_{00}=0$. However the terms describing orientation and alignment required a detailed calculation.]

\changetext{The key result of this section is Eq.~\eqref{eq:p1production}. This equation demonstrates that in the presence of a CMB quadrupole anisotropy ($a_{2,m''}$) and atoms aligned by an anisotropic local velocity gradient ($\mathscr{P}_{2,m'}$), a net magnetic moment of the atoms will develop ($\mathscr{P}_{1,m}$).} This is a new feature that is not caused by a background magnetic field (static Zeeman effect) due to its different symmetry. As we will see later, the net magnetic moment manifests itself observationally by producing circularly polarized 21 cm radiation.

\subsection{Semiclassical explanation of the effect}
\label{ss:semiclassical}

\begin{figure*}
  \centering
  \begin{subfigure}[b]{0.3\textwidth}
    \includegraphics[trim={1.5cm 2.5cm 1.5cm 0}, width=\textwidth]{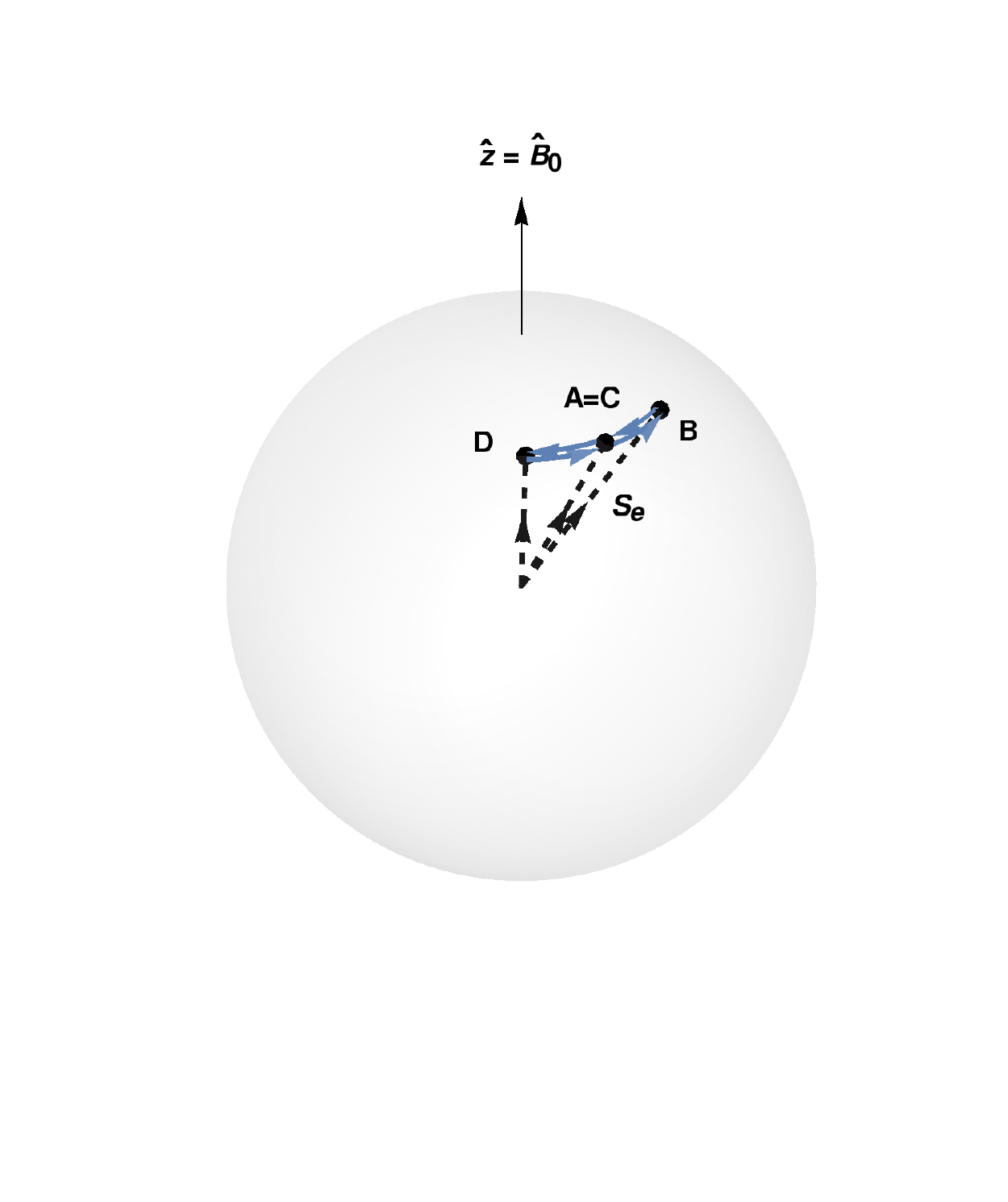}
    \caption{Isolated electron}
    \label{fig:noproton}
  \end{subfigure}
  \hfill
  \begin{subfigure}[b]{0.3\textwidth}
    \includegraphics[trim={1.5cm 2.5cm 1.5cm 0}, width=\textwidth]{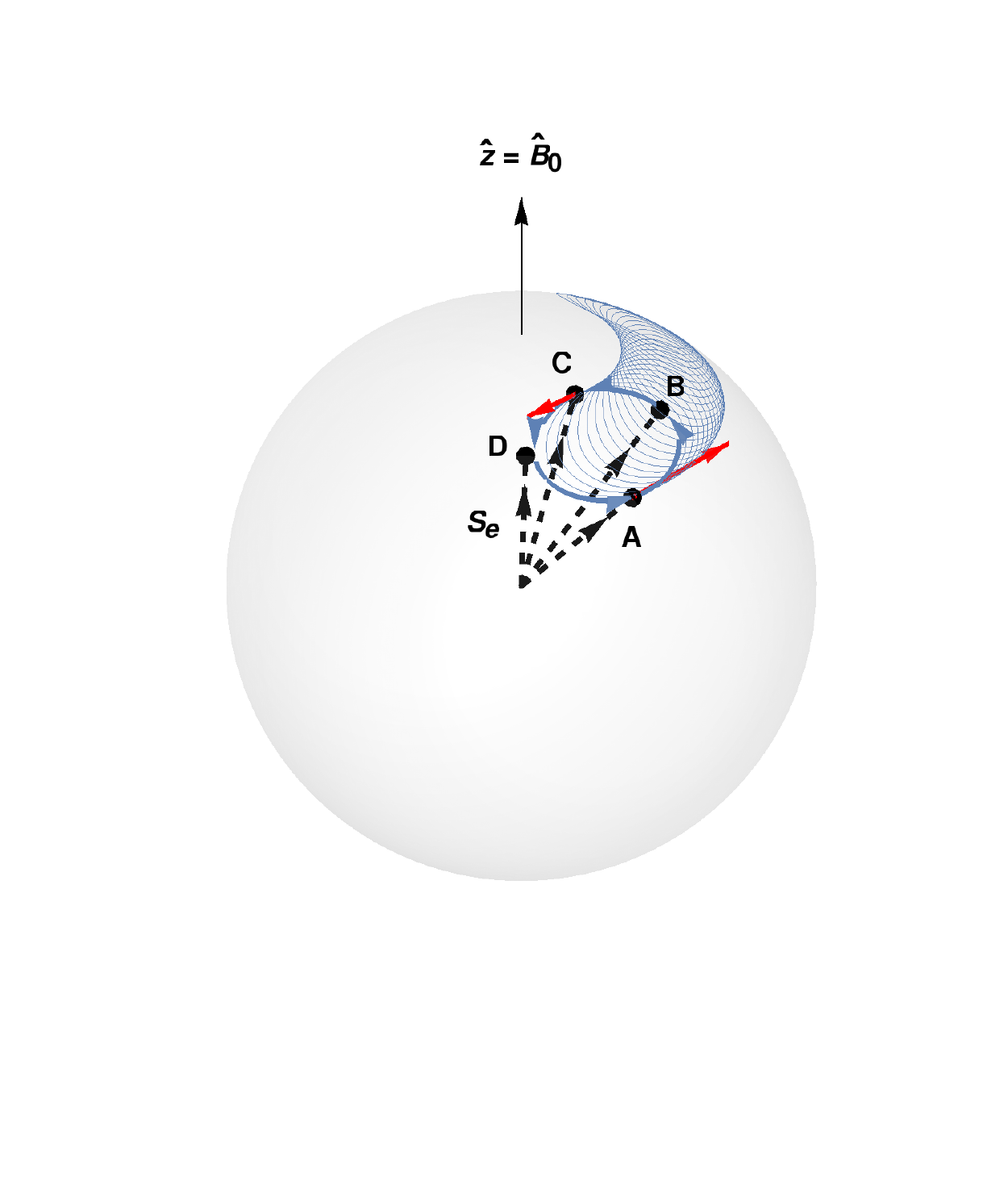}
    \caption{Electron + proton}
    \label{fig:withproton}
  \end{subfigure}
  \hfill
  \begin{subfigure}[b]{0.3\textwidth}
    \includegraphics[trim={1.5cm 2.5cm 1.5cm 0}, width=\textwidth]{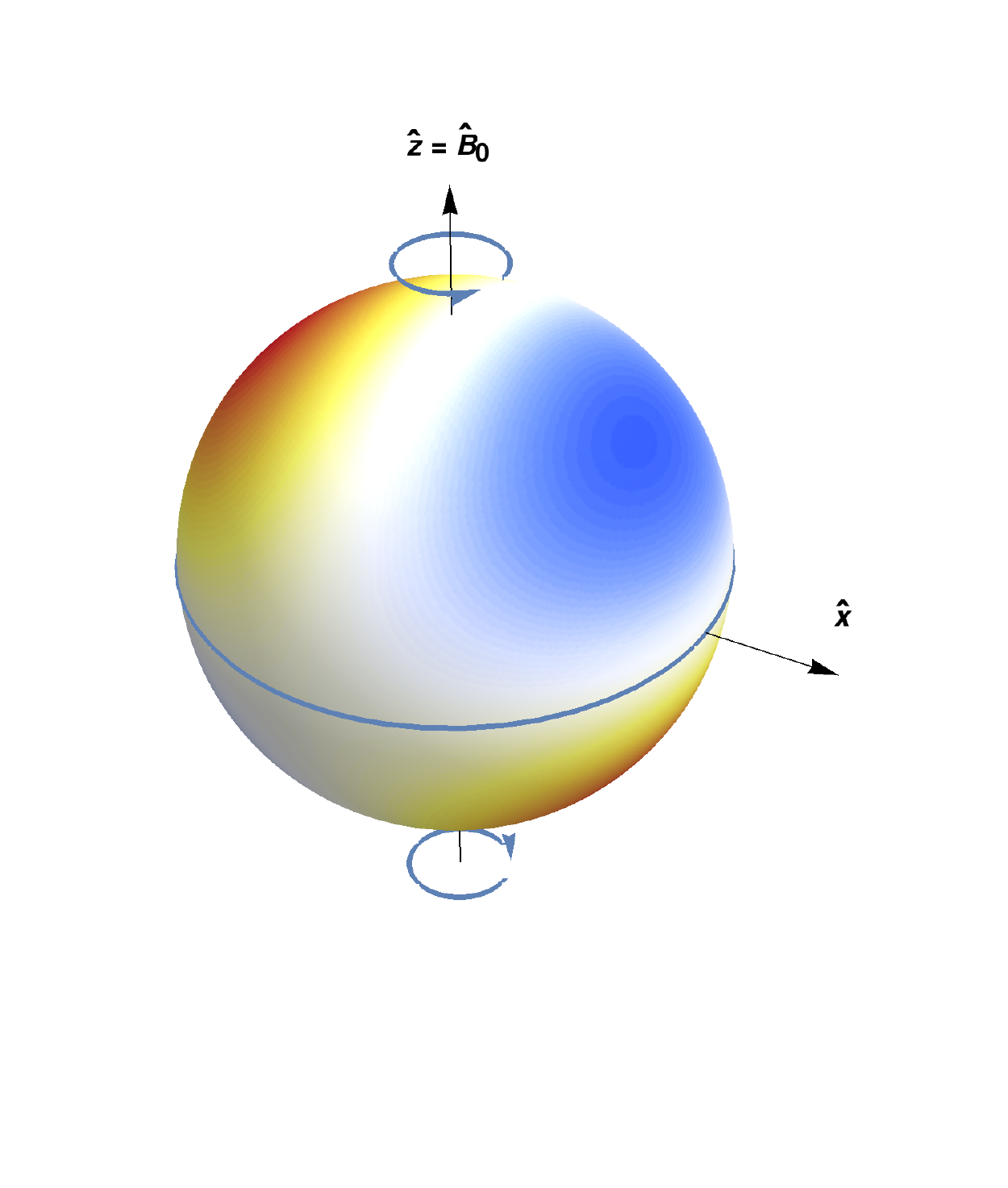}
    \caption{Effect on spin distribution}
    \label{fig:distribution}
  \end{subfigure}
  \caption{ \changetext{Semiclassical explanation of the effect. All panels assume an oscillating externally applied magnetic field, $B = B_0 \hat{z} \cos{\omega t}$. \textit{Left panel:} The blue line is the trajectory of the tip of an isolated electron's spin vector $\bm S_e$ (shown with dashed black lines). Points A, B, C, and D are positions on the trajectory at $\omega t = 0, \pi/2, \pi$, and $3\pi/2$, respectively. \textit{Middle panel:} This shows the effect of adding a proton spin $\bm S_p$ with a hyperfine interaction $(\omega_{hf}/\hbar) \bm S_e \cdot \bm S_p$. The blue trajectory and points A, B, C, and D are as in the left panel. The red lines show the instantaneous torque on the spin $\bm S_e$ due to the external magnetic field. Due to the additional oscillation along the polar direction, the average torque over a period is nonzero. Hence, the trajectory of the spin is no longer closed, and secularly drifts. \textit{Right panel:} This shows the statistical effect of the secular drift of the middle panel. Colors show a quadrupolar probability distribution (of the $x$-$z$ type; blue is higher) of the direction $\hat{\bm S}_e$ on the unit sphere. The eastward (westward) secular drift in the upper (lower) hemisphere leads to a net bias in the distribution toward $\hat{\bm y}$.} }
  \label{fig:semiclassical}
\end{figure*}

While Eq.~(\ref{eq:a-evolve}) was derived via a fully quantum mechanical calculation, it is instructive to have a semiclassical explanation of the magnetic dipole effect. In classical language, the aforementioned calculation argues that a high-frequency oscillating magnetic field (say, on the $z$-axis) can interact with atoms whose magnetic moments have no net orientation ($\langle{\bm F}\rangle=0$) but have a quadrupole alignment (say, $\langle F_xF_z\rangle>0$), and endow them with a net orientation (in this case, $\langle F_y\rangle>0$). There are some uniquely quantum aspects to this effect, but it turns out that the basic phenomenon exists in classical mechanics.

Let us consider a simple classical model consisting of an electron of spin ${\bm S}_e$ and magnetic moment ${\bm\mu}_e$, and a proton of spin ${\bm S}_p$ and magnetic moment ${\bm\mu}_p$. We assume that the electron explores a cloud around the proton with probability distribution given by the $1s$ orbital, and henceforth ignore the electron's positional degrees of freedom. We then impose an external magnetic field ${\bm B}(t)$ with zero mean value. 

\changetext{Let us first neglect the spin of the proton, i.e., assume the electron is isolated. In this case, the electron spin evolves due to the torque exerted by the external magnetic field, i.e.,
\begin{align}
 \dot{\bm S}_e &= {\bm\mu}_e \times {\bm B}(t) \nonumber \\
 &= -\frac{g_ee}{2m_ec} {\bm S}_e \times {\bm B}(t).
  \label{eq:dsei_isolated}
\end{align}
Now let us suppose that the oscillating magnetic field has the form ${\bm B}_0 \cos(\omega t)$ (we will take ${\bm B}_0$ to be in the $z$-direction, i.e., toward the North Pole in the descriptions in the text). It is possible to solve Eq.~\eqref{eq:dsei_isolated} exactly, but for our purposes, it is sufficient to treat the effect of the magnetic field perturbatively (in powers of the amplitude $B_0$) around a background state with the electron spin fixed in direction along ${\bm S}_e^{(0)}$. In this case, the first order piece ${\bm S}_e^{(1)}$ satisfies
\begin{align}
 \dot{\bm S}_e^{(1)} = -\frac{g_ee}{2m_ec} {\bm S}_e^{(0)} \times {\bm B}_0\cos(\omega t).
\end{align}
Integrating, we find
\begin{align}
  {\bm S}_e^{(1)} = -\frac{g_ee}{2m_ec\omega} {\bm S}_e^{(0)} \times {\bm B}_0\sin(\omega t). \label{eq:isolatedse}
\end{align}
We can recognize this as the standard precession of the direction of electron spin, except that the direction of precession alternates due to the oscillatory nature of the field: it is a quarter-cycle out of phase with the driving field, as expected, and is in the usual ${\bm S}_e^{(0)} \times {\bm B}_0$ direction of magnetic torque. This is illustrated in Fig.~\ref{fig:noproton}: when ${\bm B}_0$ is on the $z$-axis, the trajectory explores the ``east-west'' direction relative to the unperturbed spin vector.}

\changetext{Next, we include the spin of the proton. The torque on the electron now takes the form}
\begin{eqnarray}
\dot{\bm S}_e &=& {\bm\mu}_e \times {\bm B}(t) - \frac{\omega_{hf}}{\hbar}{\bm S}_e\times {\bm S}_p
\nonumber \\
&=& -\frac{g_ee}{2m_ec} {\bm S}_e \times {\bm B}(t) - \frac{\omega_{hf}}{\hbar}{\bm S}_e\times {\bm S}_p,
\label{eq:dse}
\end{eqnarray}
where the precession frequency of the electron spin around the proton is identified as the hyperfine splitting frequency. (The prefactor shown is quantum mechanically correct, since it corresponds to an interaction energy of $\omega_{hf}{\bm S}_e\cdot {\bm S}_p/\hbar$.) There is a similar relation for the proton,
\begin{equation}
  \dot{\bm S}_p = \frac{\omega_{hf}}{\hbar}{\bm S}_e\times {\bm S}_p. \label{eq:dsp}
\end{equation}
We neglect the direct torquing of the proton by the oscillating magnetic field, since this torque is negligible compared to that on the electron.

\changetext{We perturbatively solve Eqs.~\eqref{eq:dse} and \eqref{eq:dsp} (in powers of $B$) about a background state with the electron and proton spins parallel and fixed ${\bm S}_e^{(0)} = {\bm S}_p^{(0)}$.} We further assume that $\omega\gg\omega_{hf}$, so that the electron spin oscillates faster than it can exchange angular momentum with the proton and hence ${\bm S}_e^{(1)}\gg{\bm S}_p^{(1)}$. In this case, Eq.~(\ref{eq:dse}) gives us that
\begin{equation}
\dot{\bm S}_e^{(1)} = -\frac{g_ee}{2m_ec} {\bm S}_e^{(0)} \times {\bm B}_0\cos(\omega t)
- \frac{\omega_{hf}}{\hbar}{\bm S}_e^{(1)}\times {\bm S}_p^{(0)},
\end{equation}
or integrating:
\begin{equation}
{\bm S}_e^{(1)} = -\frac{g_ee}{2m_ec\omega} {\bm S}_e^{(0)} \times {\bm B}_0\sin(\omega t)
- \frac{\omega_{hf}}{\hbar} \left[ \int {\bm S}_e^{(1)} dt \right]\times {\bm S}_p^{(0)}.
\end{equation}
This is a recursive form for ${\bm S}_e^{(1)}$. In the limit $\omega\gg\omega_{hf}$, it has the solution
\begin{eqnarray}
\!\!\!\!\!\!\!\!{\bm S}_e^{(1)} &=& -\frac{g_ee}{2m_ec\omega} {\bm S}_e^{(0)} \times {\bm B}_0\sin(\omega t)
\nonumber \\ &&
- \frac{\omega_{hf}}{\hbar} \frac{g_ee}{2m_ec\omega^2} ({\bm S}_e^{(0)} \times {\bm B}_0)\times{\bm S}_p^{(0)} \cos(\omega t)
\nonumber \\ &&
+ {\cal O}(\omega^{-3}).
\label{eq:S1}
\end{eqnarray}
\changetext{The first term in this equation is identical to that in Eq.~\eqref{eq:isolatedse}, and has the same interpretation. The second term is new, and represents a ``north-south nodding'' in the plane containing ${\bm S}_e^{(0)}$ and ${\bm B}_0$ that is in phase with the applied field. The recursive solution for ${\bm S}_e^{(1)}$ makes its physical origin clear: the electron spin is trying to precess around the proton spin due to the hyperfine interaction. The solution is illustrated by the thick solid blue curve in Fig.~\ref{fig:withproton}. When $\omega t = -\pi/2$ (i.e., the standard precession is at its western limit; this is point D in the figure), the electron spin vector points slightly to the west of the proton spin vector, and hence the hyperfine interaction nudges the electron spin to the south and the proton spin to the north. The opposite happens at the eastern limit (point B in the figure). Hence when $\omega t = 0$ (the oscillating magnetic field points to the North Pole; point A in the figure), the electron spin is slightly south of its mean position, and when $\omega t = \pi$ (the oscillating magnetic field points to the South Pole; point C in the figure), the electron spin is slightly north of its mean position. The net result is that the spin traces out a trajectory that loops around the unperturbed direction ${\bm S}_e^{(0)}$.}

The second term in Eq.~(\ref{eq:S1}) is interesting because it is in phase with the applied magnetic field, and hence leads to a nonzero time-averaged torque. This torque is
\begin{eqnarray}
&& \!\!\!\!\!\!\!\!\!\!\!\!\!\!\!\!\!\!\!\!
\langle{\bm\mu}_e\times{\bm B}(t)\rangle
\nonumber \\
&=&  \frac{\omega_{hf}}{2\hbar \omega^2} \left(\frac{g_ee}{2m_ec}\right)^2 [({\bm S}_e^{(0)} \times {\bm B}_0)\times{\bm S}_p^{(0)}] \times {\bm B}_0
\nonumber \\
&=&  \frac{\omega_{hf}}{2\hbar \omega^2} \left(\frac{g_ee}{2m_ec}\right)^2 ({\bm B}_0\cdot{\bm S}_p^{(0)})({\bm B}_0\times{\bm S}_e^{(0)}).
\label{eq:prec}
\end{eqnarray}
This torque is eastward if the atom spin is in the northern hemisphere and westward if it is in the southern hemisphere. 

\changetext{The reason for the nonzero net torque becomes clear when we consider Fig.~\ref{fig:withproton}. Magnetic torques only act on the component of magnetic moment perpendicular to ${\bm B}$. If the atom spin is in the northern hemisphere, as in the figure, then this component ${\bm\mu}_\perp$ is larger when the electron spin is south of its mean position ($\omega t=0$, i.e. point A in the figure) and smaller when it is north of the mean position ($\omega t = \pi$, i.e. point C in the figure). Thus, there is an imbalance between eastward and westward precession, which prevents the trajectory from closing in on itself and makes it secularly drift eastward (as shown by the thin blue line in the figure). The imbalance has the opposite sign in the southern hemisphere, and the resulting drift is westward.} 

\changetext{Figure \ref{fig:distribution} shows how this secular drift acts on an initial quadrupole moment with $x$-$z$ alignment and produces a net atomic spin in the $y$-direction, $\langle F_y\rangle>0$. This is because the secular drift is inequivalent to a solid body rotation: the hot spots in the northern and hemispheres are moved eastward and westward, respectively, which biases the spin distribution towards the $\hat{\bm y}$ direction.}

Using that the electron and proton spins are of order $\hbar$ and assuming angular misalignments of order unity, the inverse timescale for this process is of order
\begin{eqnarray}
\tau^{-1}_{{\rm quad}\rightarrow{\rm dip}} &\sim&
\frac{\left| \langle{\bm\mu}_e\times{\bm B}(t)\rangle \right|}{\hbar}
\nonumber \\
&\sim& \frac{\omega_{hf}}{2\hbar \omega^2} \left(\frac{g_ee}{2m_ec}\right)^2 
\frac{B^{{\rm rad}\,2}\hbar^2}{\hbar}
\nonumber \\
&\sim& \frac{\omega_{hf}}{\omega^2} \left(\frac{g_e\mu_BB^{\rm rad}}\hbar\right)^2.
\end{eqnarray}
Using Eq.~(\ref{eq:Brms}) for the magnetic field, $\omega \sim \hbar T_\gamma/k_B$, and inserting a factor of the anisotropy $a_{20}$ since only the anisotropic part of the radiation field contributes [note in Eq.~(\ref{eq:prec}) that the torque averaged over directions of ${\bm B}_0$ vanishes], we see that this inverse timescale is indeed of order $K_{\rm mag}(1+z)^2$.

\section{Effect on 21 cm polarization}
\label{sec:poleffect}

We are now ready to compute the effect of the CMB quadrupole on the local power spectrum of 21 cm radiation and its circular polarization. We consider a small-scale Fourier mode with wave vector ${\bm k}$, in the presence of a background CMB quadrupole $a_{2m}$. \changetext{We first compute the orientation part of the density matrix $\mathscr P_{1m}$ in Sec.~\ref{ss:4a}, by taking into account precession as well as the conventional radiative processes that de-polarize the hydrogen spins. Next, in Sec.~\ref{ss:4b}, we compute the resulting circular polarization seen by a distant observer.}

\changetext{\subsection{The orientation of the hydrogen spins}
\label{ss:4a}}

\changetext{In this section, we will collect the rates of all the important processes that create or destroy the orientation part of the density matrix, $\mathscr{P}_{1m}$, and use these rates to compute the equilibrium value of the orientation.}

\changetext{In the standard picture of the 21 cm excitation, the $F=1$ hyperfine level is populated and depopulated via both collisional and radiative processes. Radiative processes, i.e., ones involving the emission or absorption of photons, are further subdivided according to whether the photons involved are resonant with the 21 cm transition. Absorption, and spontaneous and stimulated emission involve resonant photons, while the Wouthuysen-Field effect involves pumping of the hyperfine transition by nonresonant Lyman-$\alpha$ photons. The new process we study in this paper is radiative, but involves both resonant 21 cm and nonresonant CMB photons. We require the rate of change of the orientation $\mathscr{P}_{1m}$ due to all these processes.}

\changetext{First, we note that atomic collisions do not source or destroy the orientation $\mathscr{P}_{1m}$, since collisions are dominated by spin exchange and this does not affect the total (vector sum) spin of the atoms in question \cite{2017PhRvD..95h3010V}.}

\changetext{For the radiative processes, it is worth listing all the relevant quantities that are nonzero at linear order in perturbation theory. The hydrogen atom density matrix has a nonzero trace $\mathscr{P}_{00}$ (which is parametrized by the spin temperature) and alignment $\mathscr{P}_{2m}$. The 21 cm and CMB radiation fields have nonzero temperature quadrupoles (which dominate the linear polarization quadrupoles, which we neglect), but no circular polarization. None of these quantities have the right symmetry to produce orientation $\mathscr{P}_{1m}$ at linear order. Hence, at this order, radiative processes can only destroy the orientation.}

\sk{The alignment of the hydrogen spins in linear perturbation theory, caused by the anisotropic 21 cm radiation field of the surrounding gas \cite{2017PhRvD..95h3010V}, is given by
\begin{eqnarray}
\mathscr P_{2m} &=& \frac{1}{20\sqrt2\,(1+\tilde x_c+\tilde x_\alpha)}
\frac{T_\star}{T_\gamma}
\nonumber \\
&& \times \left( 1 - \frac{T_\gamma}{T_s} \right) f\delta\tau \sqrt{\frac{4\pi}5} Y_{2m}(\hat{\bm k}).
\label{eq:PI:2m}
\end{eqnarray}
Here $\tilde x_c$ and $\tilde x_\alpha$ are the collisional and Lyman-$\alpha$ coupling rates (in the convention of Ref.~\cite{2017PhRvD..95h3010V}); $f$ is the rate of growth of structure, and is $\approx 1$ in the matter-dominated era; $\delta$ is the density contrast; and $\tau$ is the 21 cm optical depth.}

\changetext{Among radiative processes involving only resonant photons, the dominant contribution is that of stimulated emission against the 21 cm background. The resulting decay of the orientation $\mathscr{P}_{1m}$ is
\begin{align}
  \dot{\mathscr P}_{1m} \vert_{\rm st. em.} & = - A \frac{T_\gamma}{T_\star} \mathscr{P}_{1m}, \label{eq:stimem}
\end{align}
where $A = 2.86 \times 10^{-15} \, {\rm s}^{-1}$ is the Einstein coefficient of the 21 cm transition, $T_\star = 68.2 \, {\rm mK}$ is the hyperfine gap in temperature units, and, as in Sec.~\ref{sec:magneticsplitting}, $T_\gamma$ is the CMB temperature.}

\changetext{The calculation of the contribution of the Wouthuysen-Field effect is more involved. We can derive this term using the methodology of \S VIC of V17. In Appendix~\ref{sec:lya}, we derive this piece by considering resonant scattering in the Lyman-$\alpha$ line. The resulting rate of decay of the alignment is
\begin{equation}
\dot {\mathscr P}_{1m}|_{{\rm Ly}\alpha} = -0.445\times6\pi \lambda_{{\rm Ly}\alpha}^2 \gamma_{2p} J(\nu_{{\rm Ly}\alpha})\,\mathscr P_{1m}, \label{eq:lya}
\end{equation}
where $\lambda_{{\rm Ly}\alpha} = 121.6 \, {\rm nm}$, $\gamma_{2p} = \Gamma_{2p}/4\pi = 50 \, {\rm MHz}$, and $J(\nu_{{\rm Ly}\alpha})$ are the wavelength, HWHM, and the input photon number flux (i.e., flux on the blue side) of the Lyman-$\alpha$ transition, respectively. In writing Eq.~\eqref{eq:lya}, we have assumed a constant photon flux across the core of the Lyman-$\alpha$ line, i.e., neglected the spectral distortion in the line itself.}

\changetext{The final piece to include is the production of alignment by the effect we propose in this paper. Note that the splitting of energies $\Delta{\cal E}_{mm'}$ of the $F=1$ level due to the CMB quadrupole, which drives the effect, is small compared to the inverse-lifetime of the state, which is $\Gamma_1 \ge A T_\gamma/T_\star$. (This formula only includes stimulated emission by the Rayleigh-Jeans tail of the CMB as a source of width; inclusion of other processes will only increase it). Therefore we treat the precession due to the CMB as a perturbation to the pre-existing alignment produced at linear order. Even though our effect enters at second order in the primordial fluctuations, it is important to include it since $\mathscr P_{1m}$ is neither present, nor produced, at linear order. We can read off this contribution from Eq.~\eqref{eq:p1production}.}

\changetext{We combine Eqs.~\eqref{eq:p1production}, \eqref{eq:stimem}, and \eqref{eq:lya}, and write the evolution equation for the orientation:
\begin{eqnarray}
\dot{\mathscr P}_{1m} &=& -\sqrt3\,iK_{\rm mag}(1+z)^2 \sum_{m'm''} \threej122{m}{m''}{-m'}
\nonumber \\ && \times
(-1)^{m'} a_{2,m''} {\mathscr P}_{2m'}
\nonumber \\ && - \frac{T_\gamma}{T_\star}A(1+\coefalpha\tilde x_\alpha) {\mathscr P}_{1m},
\label{eq:evol-1}
\end{eqnarray}
where we have rewritten the Lyman-$\alpha$ flux, $J(\nu_{{\rm Ly}\alpha})$, in terms of a dimensionless coefficient $\tilde x_\alpha$:
\begin{align}
 \tilde{x}_\alpha & = 0.445\times\frac{8\pi \lambda_{{\rm Ly}\alpha}^2 \gamma_{2p} T_\star}{A T_\gamma} J(\nu_{{\rm Ly}\alpha}) \nonumber \\
 & = 3.6 \times 10^{-2} \left( \frac{1+z}{20} \right)^{-1} \left[ \frac{J(\nu_{{\rm Ly}\alpha})}{10^{-12} {\rm cm}^{-2} {\rm Sr}^{-1} {\rm s}^{-1} {\rm Hz}^{-1}} \right],
\end{align}
which parametrizes the rate of depolarization by Lyman-$\alpha$ photons relative to that by stimulated emission.}

\sk{and the numerical prefactor is an integral over the profiles of the individual fine+hyperfine structure lines in the Lyman-$\alpha$ multiplet: 
\begin{eqnarray}
0.445 &=& \int d\nu\,\Bigl( \frac19\phi_{AA} + \frac5{27}\phi_{BB} + \frac{11}{108}\phi_{DD} + \frac5{36}\phi_{EE}
\nonumber \\
&& - \frac4{27}\phi_{AB} - \frac2{27}\phi_{AD} - \frac1{27}\phi_{BD} - \frac5{27}\phi_{BE}
\nonumber \\
&& - \frac5{54}\phi_{DE} \Bigr).
\end{eqnarray}
The Lyman-$\alpha$ flux in Eq.~(\ref{eq:evol-1}) has been re-written in terms of $\tilde x_\alpha$ instead of $J(\nu_{{\rm Ly}\alpha})$.}

\changetext{In steady-state, valid when the background parameters change on timescales long compared to the depolarization timescale $\Gamma_1^{-1} \sim T_\star/AT_\gamma$, we may set the left-hand side of Eq.~\eqref{eq:evol-1} to zero. The final value of the orientation $\mathscr{P}_{1m}$ is nonzero only to second order in the primordial fluctuations, and thus we use the linear theory value of V17 for the alignment $\mathscr{P}_{2m}$, i.e., the result in Eq.~(\ref{eq:PI:2m}). Substituting in Eq.~\eqref{eq:evol-1}, and using $T_\gamma = T_{\gamma0}(1+z)$, we find that}
\begin{eqnarray}
{\mathscr P}_{1m}
&=& -\frac{\sqrt6}{40}\,i \frac{T_\star^2 K_{\rm mag}}{T_{\gamma0}^2 A(1+\coefalpha\tilde x_\alpha)(1+\tilde x_c+\tilde x_\alpha)}
\nonumber \\
&& \times \left( 1 - \frac{T_\gamma}{T_s} \right) f\delta\tau
\sum_{m'm''} \threej122{m}{m''}{-m'}
\nonumber \\ && \times
(-1)^{m'} a_{2,m''}  \sqrt{\frac{4\pi}5}  Y_{2m'}(\hat{\bm k}).
\label{eq:POL1m}
\end{eqnarray}

\changetext{\subsection{The resulting circular polarization}
\label{ss:4b}}

\changetext{We can determine the radiation field in the vicinity of the 21 cm line in perturbation theory by repeating the analysis that lead to Eq.~(91) of V17. We describe the radiative transfer of the photons using the Boltzmann equation for the phase space density
\begin{align}
  \frac{\partial f_{\alpha \beta}}{\partial t} + c \, \hat{\bf n} \cdot \bm\nabla f_{\alpha \beta} + \frac{d\omega}{dt} \frac{\partial f_{\alpha \beta}}{d\omega} = \dot{f}_{\alpha \beta} \vert_s, \label{eq:boltzmann}
\end{align}
where $\alpha$ and $\beta$ are polarization indices. The circular polarization piece of the phase space density is defined in Eq.~\eqref{eq:fvdef}. We can isolate the $j=1$ circular polarization piece $f_{{\rm V},1m}$ in Eq.~\eqref{eq:boltzmann} using the appropriate projection in polarization space.}

\changetext{The right hand side of Eq.~\eqref{eq:boltzmann} is the source term, which describes the injection and removal of photons due to interaction with the atoms. Circular polarization is not sourced at linear order in the primordial fluctuations; from the discussion in Sec.~\ref{ss:emcpol}, we need a nonzero orientation $\mathscr{P}_{1m}$, and from Eq.~\eqref{eq:POL1m}, we see that $\mathscr{P}_{1m}$ is itself produced only at second order. Thus, the right hand side of Eq.~\eqref{eq:boltzmann} is of second order in the primordial fluctuations.}

\changetext{The left hand side in Eq.~\eqref{eq:boltzmann} describes free streaming, and the second and third terms within describe advection and redshift, respectively. As long as we restrict ourselves to cosmological fluctuations on large scales (larger than the Jeans length), we can neglect the advection term. The background and linear parts of $d\omega/dt$ and $f_{\alpha \beta}$ were calculated in V17. The linear part of $d\omega/dt$ has a quadrupole dependence on angle $\hat{\bf n}$, while at linear order the quadrupole pieces of $f_{++} + f_{--}$ and $f_{+-} = f_{-+}^\ast$ (i.e., of the intensity and linear polarization) are nonzero. These two quadrupoles are combined in the third term in Eq.~\eqref{eq:boltzmann}; however, this combination does not result in a spin-1 tensor. Thus in the final equation for the evolution of the circular polarization, we can replace $d\omega/dt$ by its background value $-H\omega$ (where $H$ is the Hubble expansion rate) and use the second order piece for $f_{\alpha \beta}/f_{{\rm V},1m}$ on the left hand side, i.e.,
\begin{align}
  \frac{\partial f_{{\rm V},1m}}{\partial t} - H \omega \frac{\partial f_{{\rm V},1m}}{d\omega} = \dot{f}_{{\rm V},1m} \vert_s. \label{eq:boltzmannv1m}
\end{align}
We now describe the evaluation of the source term on the right hand side. Among the processes listed in Sec.~\ref{ss:4a}, only the resonant processes contribute. The rates of these processes were derived in \S VII B of V17. Specifically, the rates of absorption, and spontaneous and stimulated emission are given by Eqs.~(83--85) of V17, which are phrased in terms of the moments of the photon phase space density in its `unprojected' form, i.e., $(f_{\alpha \beta})_{jm}$. We combine these equations, use Eq.~\eqref{eq:fvdef} to project out the circular polarization part, and obtain the total source term for the $j=1$ moment $f_{{\rm V},1m}$:
\begin{align}
  \dot{f}_{{\rm V}, 1 m} (\omega) \vert_s & = n_{\rm H} x_{\rm 1s} \frac{\sigma(\omega) c}{3} \biggl[ -\left(3 - 4 \mathscr{P}_{0 0} \right) f_{{\rm V}, 1 m} (\omega) \notag \\
  & ~~~ + \sqrt{\frac{3}{2}} \left(1 + f_{{\rm I}, 0 0}\right) \mathscr{P}_{1 m} \biggr], \label{eq:v1msource}
\end{align}
where $n_{\rm H}$ is the hydrogen number density, $x_{\rm 1s}$ is the neutral fraction, and $\sigma(\omega)$ is the absorption cross-section for the 21 cm line. In writing Eq.~\eqref{eq:v1msource}, we have neglected stimulation emission involving the moments $f_{{\rm I}, 2m}$, $f_{{\rm V}, 3m}$, and $\mathscr{P}_{2m}$, since these latter terms are themselves of higher order in the 21 cm optical depth.}

\changetext{Next, we substitute Eq.~\eqref{eq:v1msource} into Eq.~\eqref{eq:boltzmannv1m}, and drop the time derivative $\partial f/\partial t$ in the vicinity of the line. This is equivalent to assuming that a steady state develops, with the injection of photons by radiative processes balanced by the redshifting due to Hubble expansion. Under this assumption, the radiative transfer equation reduces to}
\begin{equation}
\frac{\partial f_{{\rm V},1m}}{\partial \cal X} = \tau \left[ f_{{\rm V},1m} - \sqrt{\frac83} \frac{T_\gamma T_s}{T_\star^2} \mathscr P_{1m} \right],
\end{equation}
\changetext{where ${\cal X}$ is the cumulative line profile (ranging from 0 at the red edge of the line to 1 at the blue edge), the factors involving the absorption cross-section give the 21 cm optical depth $\tau$, and in the final term in Eq.~\eqref{eq:v1msource}, the spontaneous emission has been neglected relative to the stimulated emission by the phase-space density $f_{{\rm I},00} \approx T_\gamma/T_*$.} With the boundary condition of no ``input'' circular polarization (i.e. $f_{{\rm V},1m}=0$ at ${\cal X}=1$) and in the limit of $\tau\ll 1$, the solution at the red edge ${\cal X}=0$ is
\begin{equation}
f_{{\rm V},1m}({\cal X}=0) = \sqrt{\frac83} \frac{T_\gamma T_s}{T_\star^2} \tau \mathscr P_{1m}.
\label{eq:POL-f1m}
\end{equation}
An observer looking in the $+z$ direction (i.e. looking at photons propagating in the $-z$ direction) sees a phase space density in circular polarization of $-f_{V,10}$, or in temperature units
\begin{equation}
V_{\rm obs} = -\frac{T_\star}{1+z}f_{{\rm V},10}.
\end{equation}
Putting this together with Eqs.~(\ref{eq:POL1m}) and (\ref{eq:POL-f1m}) yields
\begin{eqnarray}
V_{\rm obs} &=& \frac{\sqrt\pi}{5\sqrt{5}}\,i\,\frac{T_s T_\star K_{\rm mag}  f \tau^2\delta}{T_{\gamma 0}A (1+\coefalpha\tilde x_\alpha)(1+\tilde x_c+\tilde x_\alpha)}
\nonumber \\
&& \times \left( 1 - \frac{T_\gamma}{T_s} \right)
\sum_{m'm''} (-1)^{m'} \threej1220{m''}{-m'}
\nonumber \\ &&
\times a_{2,m''}   Y_{2m'}(\hat{\bm k}).
\end{eqnarray}
Substituting the $3j$ symbols allows us to expand the sum. There are only nonzero terms for $m''=m'$ and $m'\neq 0$. The terms with $m'\leftrightarrow-m'$ are negative complex conjugates of each other, which allows us to write the sum in terms of the imaginary part of only 2 terms. This leads to
\begin{eqnarray}
\label{eq:vobs}
V_{\rm obs} &=& -\frac{\sqrt{2\pi}}{25\sqrt{3}}\,\frac{T_s T_\star K_{\rm mag}  f \tau^2\delta}{T_{\gamma 0}A (1+\coefalpha\tilde x_\alpha)(1+\tilde x_c+\tilde x_\alpha)}
\nonumber \\
&& \times \left( 1 - \frac{T_\gamma}{T_s} \right)
\Im [
a_{21} Y_{21}(\hat{\bm k})
+2a_{22} Y_{22}(\hat{\bm k})
].
\nonumber \\ &&
\end{eqnarray}
Thus, we see that the circular polarization transfer function $\partial V_{\rm ons}/\partial\delta$ depends on the direction of the wavenumber $\hat{\bm k}$.

For standard cosmological parameters (as in \changetext{V17}), with a mean optical depth of $\tau = 0.0097(T_\gamma/T_s)[(1+z)/10]^{1/2}$, and with $f=1$ in the matter-dominated era, this transfer function takes the numerical value
\begin{eqnarray}
\label{eq:transfer}
\frac{\partial V_{\rm obs}}{\partial \delta} &=& -8.6\,{\rm mK} \left(\frac{1+z}{20}\right)^{2} \frac{T_\gamma}{T_s}\left( 1 - \frac{T_\gamma}{T_s} \right)
\nonumber \\
&& \times \frac1{(1+\coefalpha\tilde x_\alpha)(1+\tilde x_c+\tilde x_\alpha)}
\nonumber \\ && \times 
\Im [a_{21} Y_{21}(\hat{\bm k})
+2a_{22} Y_{22}(\hat{\bm k})
].
\end{eqnarray}
The transfer function (and by extension the associated circular polarization power spectra) thus depend on 4 of the 5 types of the CMB quadrupole moments. The circular polarization signal does not depend on the $m=0$ CMB quadrupole mode that is symmetric around the line of sight.

\section{Summary and Discussion}
\label{sec:discussion}

In this paper we have shown that the cosmological 21 cm radiation should pick up a small circular polarization due to the quadrupole moment of the CMB. The signal is very small; for typical CMB quadrupole moments of $a_{2m}\sim 10^{-5}$, Eq.~(\ref{eq:transfer}) predicts a circular polarization of $\sim 0.1\,\mu$K times the density perturbation $\delta$. This is five orders of magnitude fainter than the intensity signal that is the target of current experiments. Nevertheless, the signature is very different from other 21 cm signals discussed in the literature. We thus propose it as a method to measure the remote quadrupole of the CMB during the cosmic Dark Ages.

The physical basis of this method relies on the splitting of the $F = 1$ hyperfine level of neutral hydrogen due to the remote (i.e.\ at the position of the emitting gas, rather than the observer) quadrupole moment of the CMB. Unlike the Zeeman effect, where $M_{F}=\pm 1$ have opposite energy shifts, the remote CMB quadrupole shifts $M_{F}=\pm 1$ together relative to $M_{F}= 0$. This splitting leads to a small circular polarization of the emitted 21cm photon, which encodes information about the remote CMB quadrupole through Eq.~(\ref{eq:vobs}). The calculation assumes that the magnetic field is small compared to the saturation value, which would have to be verified from the 21 cm intensity power spectrum \cite{2017PhRvD..95h3010V,2017PhRvD..95h3011G}. If this assumption is valid, it would also ensure that there is no circular polarization resulting from the Zeeman-induced radial displacement of the 21 cm-based density maps in the right vs.\ left circular polarizations.

To estimate the circular polarization signal we present a detailed calculation of the atomic density matrix coupled to an  anisotropic non-resonant photon bath. This yields the relative change in the sub level energies ${\cal E}_{M_F=1}-{\cal E}_{M_F=0}$. The main results of this paper are Eqs.(\ref{eq:vobs}) and (\ref{eq:transfer}) which show that the circular polarization signal depends on four of the five types of quadrupole moments of the CMB. While the CMB quadrupole and atomic physics determine the angular structure of the signal, the amplitude depends on astrophysical inputs such as the gas temperature and Lyman-$\alpha$ flux.

The method outlined in this paper presents a novel method to construct a {\it remote CMB quadrupole} field, using the circular polarization of the redshifted 21 cm line. Such a field can in turn be decomposed into $E$ and $B$-modes, much like the CMB polarization field; just as for the CMB polarization, it turns out that the $B$-mode piece can be generated by tensor modes but not scalar modes. In Paper II of this series (Mishra \& Hirata 2017), we discuss the detectability of the 21 cm circular polarization signal, and present forecasts for measuring the the $B$-modes of the remote quadrupole field and predicted uncertainties on $r$ with future radio arrays.

\begin{acknowledgments}
We thank Vera Gluscevic and Antonija Oklop\v ci\'c for enlightening conversations and comments during the preparation of this work.
CH and AM are supported by the U.S. Department of Energy and the David \& Lucile Packard Foundation. CH is supported by the Simons Foundation and the National Aeronautics and Space Administration.
TV gratefully acknowledges support from the
Schmidt Fellowship and the Fund for Memberships in
Natural Sciences at the Institute for Advanced Study.
\end{acknowledgments}

\appendix

\section{Electric dipole splitting}
\label{sec:ed}

We now compute the splitting of the $F=1$ level of hydrogen by the electric field of the anisotropic CMB (dynamic Stark effect). It is shown herein that the splitting is negligible compared to that of the magnetic dipole splitting. In the main text, the electric dipole splitting is therefore ignored.

Since the interaction energy for an electric dipole is $-{\bm d}\cdot{\bm E}$, the electric dipole energy shift is very similar in form to the magnetic dipole energy shift, Eq.~(\ref{eq:ESm.d.}):
\begin{eqnarray}
\!\!\!\!\!\!\!\!
\Delta {\cal E}_{ji}^{\rm e.d.} &=&
\sum_{n\mu\nu} \langle :\!E^{{\rm rad}(-)}_\mu E^{{\rm rad}(+)}_\nu\!: \rangle
\nonumber \\ && \times
\left[
\frac{(d_\mu)_{ni}(d_\nu)_{jn} }{{\cal E}_j-{\cal E}_{n}-\hbar\omega}
+ \frac{(d_\nu)_{ni}(d_\mu)_{jn}}{{\cal E}_j-{\cal E}_{n}+\hbar\omega}
\right].
\label{eq:ESe.d.}
\end{eqnarray}
The principal difference is that the electric dipole operator connects $1s_{1/2}(F=1)$ to the $np_j(F)$ states. Since the electric dipole operator does not act on the spin state of the electron or proton, if there were no fine or hyperfine structure in the excited states, then we could choose a basis of definite quantum numbers $nlm_lQM_Q$ (where the total spin angular momentum ${\bm Q}={\bm S}+{\bm I}$ excludes the orbital angular momentum; i.e. the total angular momentum is ${\bm F} = {\bm L}+{\bm Q}$), and then the expression in Eq.~(\ref{eq:ESe.d.}) would be trivially diagonal in $QM_Q$ and independent of $QM_Q$. Therefore it is profitable to separately break out the non relativistic Hamiltonian ($H^{(0)}$) and the perturbation $H^{(1)}$ (which includes fine and hyperfine structure). The first term in brackets in Eq.~(\ref{eq:ESe.d.}), summed over the intermediate state, can be represented as
\begin{equation}
\sum_n\frac{(d_\mu)_{ni}(d_\nu)_{jn} }{{\cal E}_j-{\cal E}_{n}-\hbar\omega}
=
\langle j | d_\nu \frac1{E_j - H - \hbar\omega} d_\mu | i\rangle,
\label{eq:ESji}
\end{equation}
where $j$ and $i$ are the true initial and final states. If we split $H = H^{(0)}+H^{(1)}$, and suppose that $|i\rangle$ is the eigenstate of $H$ corresponding to the unperturbed eigenstate $|1s,Q'M_{Q'}\rangle$ of $H^{(0)}$ and $|j\rangle$ is the eigenstate of $H$ corresponding to the unperturbed eigenstate $|1s,QM_Q\rangle$, then this matrix element of Eq.~(\ref{eq:ESji}) is
\begin{widetext}
\begin{eqnarray}
&&\!\!\!\!\!\!\!\!
\langle 1s,QM_Q| d_\nu \frac1{E_j - H^{(0)} - \hbar\omega} d_\mu | 1s,Q'M_{Q'}\rangle
+ \langle 1s,QM_Q| d_\nu \frac1{E_j - H^{(0)} - \hbar\omega} H^{(1)} \frac1{E_j - H^{(0)} - \hbar\omega} d_\mu | 1s,Q'M_{Q'}\rangle
\nonumber \\ &&
+ \langle 1s,QM_Q| H^{(1)}\Pi \frac1{E_j - H^{(0)}} d_\nu \frac1{E_j - H^{(0)} - \hbar\omega} d_\mu | 1s,Q'M_{Q'}\rangle
\nonumber \\ &&
+ \langle 1s,QM_Q| d_\nu \frac1{E_j - H^{(0)} - \hbar\omega} d_\mu \frac1{E_j-H^{(0)}} \Pi H^{(1)} | 1s,Q'M_{Q'}\rangle.
\label{eq:ES-temp1}
\end{eqnarray}
This is to first order in $H^{(1)}$ and including both the perturbations to the operator $H$ and to the initial and final eigenstates. The operator $\Pi$ projects out the original state $|1s,QM_{Q}\rangle$ or $|1s,Q'M_{Q'}\rangle$ (in general, it may be taken as a projection that removes the unperturbed $1s$ states). Of the 4 terms in Eq.~(\ref{eq:ES-temp1}), the first one is proportional to $\delta_{QQ'}\delta_{M_QM_{Q'}}$ -- i.e. it produces the same energy shift for all of the states in the $1s$ configuration. Thus it can be neglected for the purposes of obtaining energy splittings. We thus consider the contributions involving the perturbation $H^{(1)}$.

The perturbation $H^{(1)}$ contains relativistic terms that are spin-independent and hence of no interest to us, fine structure terms $\propto {\bm L}\cdot{\bm S}/r^3$, and hyperfine structure (interaction of the electron and proton magnetic moments). We first consider the fine structure terms, since they are larger than hyperfine structure by the ratio of the electron to proton magnetic moment ($\mu_e/\mu_p\sim 10^3$). Since the $1s$ configuration states are eigenstates of ${\bm L}$ with eigenvalue zero, the fine structure Hamiltonian $H^{(1)}_{fs}$ contributes only to the second term in Eq.~(\ref{eq:ES-temp1}). This term is however antisymmetric in $\mu$ and $\nu$ since a term $\propto {\bm L}\cdot{\bm S}/r^3$ can be factored as
\begin{eqnarray}
&& \!\!\!\!\!\!\!\!\!\!\!\!\!\!\!\!\!\!\!\!\!\!\!\!
\sum_\sigma
\langle 1s| d_\nu \frac1{E_j - H^{(0)} - \hbar\omega} \frac{L_\sigma}{r^3} \frac1{E_j - H^{(0)} - \hbar\omega} d_\mu | 1s\rangle
\langle QM_Q|S_\sigma|Q'M_{Q'}\rangle
\nonumber \\
&=&
\sum_\sigma
\langle 1s| d_\nu \frac1{E_j - H^{(0)} - \hbar\omega} r^{-3} \frac1{E_j - H^{(0)} - \hbar\omega} L_\sigma d_\mu | 1s\rangle
\langle QM_Q|S_\sigma|Q'M_{Q'}\rangle
\nonumber \\
&=& i\hbar
\sum_\sigma \epsilon_{\sigma \mu \rho}
\langle 1s| d_\nu \frac1{E_j - H^{(0)} - \hbar\omega} r^{-3} \frac1{E_j - H^{(0)} - \hbar\omega}  d_\rho | 1s\rangle
\langle QM_Q|S_\sigma|Q'M_{Q'}\rangle
\nonumber \\
&=& i\hbar
\sum_\sigma \epsilon_{\sigma \mu \nu} C_E
\langle QM_Q|S_\sigma|Q'M_{Q'}\rangle
,
\end{eqnarray}
\end{widetext}
where in the first equality we used that $[H^{(0)},L_\sigma]=0$ since $H^{(0)}$ is rotationally invariant and contains no spins; in the second equality we used that the dipole moment operator is a vector so that $[L_\sigma, d_\mu] = i\hbar \epsilon_{\sigma \mu \rho} d_\rho$ and that $L_\sigma|1s\rangle =0$; and finally, since the operators $H^{(0)}$ and $r^{-3}$ are spherically symmetric, the spatial matrix element in the third line must have the form $C_E \delta_{\nu\rho}$, where $C_E$ is some constant. The $\mu\nu$ antisymmetry of the resulting expression implies that it only couples to circularly polarization of the incident radiation field (see Eq.~\ref{eq:ESe.d.}). Since the circular polarization of the CMB is a factor of $\ll 10^{-3}$ smaller than the anisotropies, we will neglect it unless the hyperfine Hamiltonian contributions that come from anisotropic radiation are suppressed by some symmetry.

The hyperfine Hamiltonian given by Eq.~(22.1) of Ref.~\cite{1957qmot.book.....B} contains terms of the form
\begin{equation}
H^{(1)}_{hf} \ni \frac{2g_p\mu_N\mu_B}{\hbar^2 r^3}(3\hat r_\alpha \hat r_\beta - \delta_{\alpha\beta})S_\alpha I_\beta,
\label{eq:ES-hf}
\end{equation}
where ${\mathbf r}$ is the electron position operator.
(The remaining terms either contain factors of ${\bm L}\cdot{\bm I}$ or ${\bm S}\cdot{\bm I}$; the former is suppressed since again it couples only to circular polarization, and the latter is not relevant for energy splittings since for all $1s$ $F=1$ states, ${\bm S}\cdot{\bm I}=\frac14\hbar^2$ is a constant.) The term in Eq.~(\ref{eq:ES-hf}), however, leads to a nonzero contribution in Eq.~(\ref{eq:ES-temp1}). In the limit where $\omega\ll\omega_{{\rm Ly}\alpha}$ so that we can neglect $\hbar\omega$ in the denominators in Eq.~(\ref{eq:ES-temp1}), we find that Eq.~(\ref{eq:ES-temp1}) reduces to
\begin{eqnarray}
\!\!\!\!\!\!\!\!\!\!
\sum_n\frac{(d_\mu)_{ni}(d_\nu)_{jn} }{{\cal E}_j-{\cal E}_{n}}
\!\! &=& \!\! 2e^2g_p\mu_N\mu_B {\cal K}_{\nu\mu,\alpha\beta}
\nonumber \\ && \times
 \langle 1M_F(j)|\frac{S_\alpha I_\beta}{\hbar^2} |1M_F(i)\rangle,
\end{eqnarray}
where the constant ${\cal K}_{\nu\mu,\alpha\beta}$ is given by
\begin{eqnarray}
{\cal K}_{\nu\mu,\alpha\beta} \!\! &=& \!\! \langle 1s| (
r_\nu G q_{\alpha\beta} G r_\mu + q_{\alpha\beta} \Pi G r_\nu G r_\mu
\nonumber \\ && + r_\nu G r_\mu G \Pi q_{\alpha\beta}
)|1s\rangle
\end{eqnarray}
and we have used that $d_\mu=-er_\mu$, introduced the quadrupole position operator $q_{\alpha\beta} = (3\hat r_\alpha \hat r_\beta - \delta_{\alpha\beta})/r^3$ and the Green's function $G = (E_{1s}-H^{(0)})^{-1}$. (Technically, $G$ should use the energy of the true hyperfine-split level $E_{F=1}$ rather than the unperturbed $1s$ energy. This ``residual'' correction \cite{1968RSPSA.305..125A} has no effect on the splitting of the $M_F$ sub levels and is ignored here.) The projector $\Pi$ is presented here but is technically unnecessary since it always acts on a state with $l=2$. The $\mu\rightarrow\nu$ symmetry and the traceless-symmetric nature of $q_{\alpha\beta}$ force ${\cal K}_{\nu\mu,\alpha\beta}$ to have the form
\begin{equation}
{\cal K}_{\nu\mu,\alpha\beta} = K \left(\frac12 \delta_{\nu\alpha} \delta_{\mu\beta} + \frac12 \delta_{\nu\beta} \delta_{\mu\alpha}
- \frac13\delta_{\mu\nu}\delta_{\alpha\beta} \right),
\end{equation}
where $K$ is a constant. A numerical evaluation gives
\begin{equation}
K = 2.350 \frac{a_0}{e^4};
\end{equation}
the computation in Ref.~\cite{1967PPS....92..857S} translated into the language of our discussion gives the exact analytic value of the pre factor as $\frac{47}{20}$, in agreement with our numerical estimate. (An earlier version of the calculation is given by Ref.~\cite{1959AnPhy...6..156S}, although it appears to be missing some terms.)
It then follows that the radiation-induced energy splitting in the $F=1$ sub levels is:
\begin{eqnarray}
\!\!\!\!\!\!\!\!
\Delta {\cal E}_{ji}^{\rm e.d.} &=&
4e^2g_p\mu_N\mu_B K \sum_{\mu\nu} \langle :\!E^{{\rm rad}(-)}_\mu E^{{\rm rad}(+)}_\nu\!: \rangle
\nonumber \\ && \times
 \langle 1M_F(j)|\frac{S_{\langle\mu} I_{\nu\rangle}}{\hbar^2} |1M_F(i)\rangle,
\label{eq:ES-delta}
\end{eqnarray}
where the angle brackets refer to the traceless-symmetrization of the indices shown.

The shift in Eq.~(\ref{eq:ES-delta}) is sourced entirely by the traceless-symmetric part of the electric field covariance, which is in turn sourced by the CMB temperature quadrupole anisotropy (and the polarization quadrupole, but we ignore this here since the temperature anisotropy is larger). In the case of isotropic radiation, the electric field covariance is
\begin{equation}
\langle :\!E^{{\rm rad}(-)}_\mu E^{{\rm rad}(+)}_\nu\!: \rangle_{\rm isotropic} = \frac23\pi a_{\rm rad}T_\gamma^4
\delta_{\mu\nu}.
\end{equation}
If there is a radiation quadrupole $a_{20}$, then using the rules derived in \S\ref{ss:md} for the temperature seen in different directions, the traceless-symmetric part of this is
\begin{equation}
\langle :\!E^{{\rm rad}(-)}_{\langle\mu} E^{{\rm rad}(+)}_{\nu\rangle}\!: \rangle = -\frac{4\sqrt\pi}{3\sqrt5} a_{\rm rad}T_\gamma^4 a_{20}
\left( \begin{array}{ccc} -\frac12 & 0 & 0 \\ 0 & -\frac12 & 0 \\ 0 & 0 & 1 \end{array} \right).
\end{equation}
Since this quadrupole is symmetric around the $z$-axis, it does not mix different $M_F$ states, but it does mean that the $M_F=\pm1$ states shift relative to the $M_F=0$ state. Using the matrix elements
\begin{equation}
\langle 11|\frac{S_{\langle\mu} I_{\nu\rangle}}{\hbar^2} |11\rangle
- \langle 10|\frac{S_{\langle\mu} I_{\nu\rangle}}{\hbar^2} |10\rangle
= \left( \begin{array}{ccc} -\frac14 & 0 & 0 \\ 0 & -\frac14 & 0 \\ 0 & 0 & \frac12 \end{array}\right),
\end{equation}
we see that
\begin{eqnarray}
\!\!\!\!\!\!\!\!\!\!\!\!
\Delta {\cal E}_{11}^{\rm e.d.} - \Delta {\cal E}_{10}^{\rm e.d.} &=& -\frac{4\sqrt\pi}{\sqrt5}
e^2g_p\mu_N\mu_B K  a_{\rm rad}T_\gamma^4 a_{20}
\nonumber \\
&=& -4\times10^{-12}\,{\rm s}^{-1} \left( \frac{T_\gamma}{60\,\rm K} \right)^4 a_{20}.
\label{eq:ans-ed}
\end{eqnarray}
This effect is 2---3 orders of magnitude smaller than the magnetic dipole effect, and so it is neglected here.

We note that the blackbody radiation-induced shift in the $^{133}$Cs hyperfine transition frequency is due mainly to the electric dipole rather than the magnetic dipole effect \cite{2006PhRvL..97d0802A, 2006PhRvL..97d0801B}. The difference relative to the case of the hydrogen atom is two-fold: (i) the existence of low-lying electric dipole transitions in the alkalis (e.g. [Xe]$6s\rightarrow\,$[Xe]$6p$ at 1.4 eV in Cs versus $1s\rightarrow 2p$ at 10.2 eV in H) with large matrix elements strongly enhances the Stark effect; and (ii) the bulk of the quadratic Stark shift in the $^1$H hyperfine frequency involves the contact interaction $\propto {\bm S}\cdot{\bm I} \delta^{(3)}({\bm r})$ \cite{1967PPS....92..857S}, which does not lift the degeneracy among the $M_F$-sublevels.

\section{Splittings from other radiation sources}
\label{sec:other}

We have considered the splitting of $M_F$-sublevels due to the CMB anisotropy. However, in principle we must consider splittings from other sources of anisotropic radiation. These sources are weaker than the CMB but may have larger quadrupole moments. The three considered here are the 21 cm background itself, the kinematic quadrupole, and (at lower redshift) starlight from early sources.

\subsection{The 21 cm anisotropy}
\label{ss:21cm}

To study the effect of the 21 cm anisotropy, we must return to Eq.~(\ref{eq:MDF1}) because this background consists of radiation at $\omega \sim {\cal O}(\omega_{hf})$ and hence the high-frequency limit used for the CMB is inapplicable. We may instead replace the blackbody formula for the magnetic field fluctuations with the Rayleigh-Jeans limit,
\begin{equation}
\langle :\! B_\mu^{{\rm rad}(-)} B_\nu^{{\rm rad}(+)} \!:\rangle = \frac{2k_B}{3\pi c^3} \delta_{\mu\nu} \int \omega^2 T_{\rm RJ}(\omega)\,d\omega,
\end{equation}
where $T_{\rm RJ}(\omega)$ is the classical (Rayleigh-Jeans) radiation temperature.
The replacement for Eq.~(\ref{eq:MDe}) in the case of a quadrupole moment in the $a_{20}$ mode in the long-wavelength radiation is then
\begin{eqnarray}
\frac{\Delta{\cal E}^{\rm m.d.}_{11}-\Delta{\cal E}^{\rm m.d.}_{10}}{\hbar} &=&
\frac{k_B\mu_B^2\omega_{hf}}{\sqrt5\,\pi^{3/2} \hbar^2 c^3}
\int \frac{\omega^2}{\omega^2-\omega_{hf}^2} 
\nonumber \\ && \times
T_{\rm RJ}(\omega)a_{20}(\omega)\,d\omega.
\end{eqnarray}
For radiation sources such as the 21 cm radiation that have $\omega \sim {\cal O}(\omega_{hf})$,
the integral will be dominated by the regime where $\omega\approx\omega_{hf}$. In this case, we may approximate $\omega^2/(\omega^2-\omega_{hf}^2) \approx \omega_{hf}/2(\omega-\omega_{hf})$. Endowing $\omega_{hf}$ with an infinitesimal positive imaginary part (equivalent to giving the $|1s,F=0\rangle$ state an exponentially decaying natural amplitude) then gives
\begin{equation}
\frac{\omega^2}{\omega^2-\omega_{hf}^2} \approx {\mathfrak P} \frac{\omega_{hf}}{2(\omega-\omega_{hf})} 
+ \frac{i\pi\omega_{hf}}{2} \delta(\omega-\omega_{hf}),
\end{equation}
where ${\mathfrak P}$ denotes the principal part (significant only when taking the integral over $\omega$). This results in
\begin{eqnarray}
\frac{\Delta{\cal E}^{\rm m.d.}_{11}-\Delta{\cal E}^{\rm m.d.}_{10}}{\hbar} &=&
\frac{k_B\mu_B^2\omega_{hf}^2}{2\sqrt5\,\pi^{3/2} \hbar^2 c^3} \biggl[
{\mathfrak P}\!\!\int \frac{T_{\rm RJ}a_{20}(\omega)}{\omega-\omega_{hf}}\,d\omega
\nonumber \\ &&
+ i\pi T_{\rm RJ}(\omega_{hf}) a_{20}(\omega_{hf})
\biggr].
\end{eqnarray}
Using that the Einstein coefficient for the 21 cm line is $A = 4\mu_B^2(\omega_{hf}/c)^3/(3\hbar)$ and that $T_\star = \hbar\omega_{hf}/k_B$, this simplifies to
\begin{eqnarray}
\frac{\Delta{\cal E}^{\rm m.d.}_{11}-\Delta{\cal E}^{\rm m.d.}_{10}}{\hbar} &=&
\frac{3A}{\sqrt{320\pi}\, T_\star} \biggl[ \frac1\pi
{\mathfrak P}\!\!\int \frac{T_{\rm RJ}a_{20}(\omega)}{\omega-\omega_{hf}}\,d\omega
\nonumber \\ &&
+ i T_{\rm RJ}(\omega_{hf}) a_{20}(\omega_{hf})
\biggr].
\label{eq:.Q1}
\end{eqnarray}
Since the difference of the temperatures seen by the $x$ and $y$ dipoles and the $z$ dipole is $3/\sqrt{80\pi}\,T_{\rm RJ}a_{20}(\omega_{hf})$, the last term can be identified as the difference in lifetime (imaginary energy) due to the orientation-dependent probability for stimulated emission. This effect is already taken into account in the formalism of \changetext{V17} and should not be double-counted; it is therefore dropped here.

In order to establish whether the splitting of the hyperfine line by ambient 21 cm radiation is significant, we need an order-of-magnitude argument for Eq.~(\ref{eq:.Q1}). This can be obtained by supposing that at any given point ${\bm r}$, the 21 cm radiation from neighboring points is of the form
\begin{equation}
T_{\rm RJ}({\bm r},\omega,\hat{\bm n}) = [{\rm isotropic}] + \frac{\partial T_{21}}{\partial \delta} \delta({\bm r} + s\hat{\bm n}),
\end{equation}
where $s = (\omega_{hf}-\omega)/(aH)$ is the comoving distance over which a photon redshifts from the hyperfine frequency $\omega_{hf}$ to $\omega$. The multiplying factor $\partial T_{21}/\partial\delta$ (units of K) is the change in 21 cm brightness temperature per unit change in the over density, measured at the redshift of interest (i.e. it is a factor of $1+z$ greater than the corresponding factor observed at Earth and thus reported in predictions of the 21 cm signal \cite{2004PhRvL..92u1301L}). This approach neglects redshift-space distortions, which should suffice for an order of magnitude calculation. We take $T_{\rm RJ}$ to be isotropic blue ward of the 21 cm line, since the 21 cm emission (or absorption) has no effect there. We then find that the frequency splitting in Eq.~(\ref{eq:.Q1}) reduces to
\begin{equation}
\Delta\omega_{10} =
\frac{3A}{\sqrt{320\pi^3}\, T_\star} \frac{\partial T_{21}}{\partial\delta}
\int \delta({\bm r} + s\hat{\bm n}) Y_{20}^\ast({\bm n}) \frac{ds}s d^2\hat{\bm n}.
\label{eq:.Q2}
\end{equation}
The variance of the integral ${\cal I}$ can be obtained from the power spectrum of the matter,
\begin{eqnarray}
{\rm Var\,}{\cal I} &=& \int \frac{d^3\bm k}{(2\pi)^3}\,P_\delta(k)
\left| \int d^2\hat{\bm n}\,Y_{20}^\ast(\hat{\bf n}) \int \frac{ds}s \,e^{i{\bm k}\cdot s\hat{\bm n}}  \right|^2
\nonumber \\
&=& \int \frac{d^3\bm k}{(2\pi)^3}\,P_\delta(k)
\left| -4\pi\,Y_{20}^\ast(\hat{\bf k}) \int \frac{ds}s \,j_2(ks) \right|^2
\nonumber \\
&=& \int \frac{2k^2\,dk}{9\pi}\,P_\delta(k)
\nonumber \\
&=& \frac{4\pi}9 \sigma^2_\delta,
\end{eqnarray}
where we have used the identity $\int_0^\infty j_2(x)\,dx/x = \frac13$. Thus the root-mean-square frequency splitting coming from the $20$ quadrupole moment is
\begin{equation}
[{\rm Var\,}\Delta\omega_{10}]^{1/2} = \frac{A}{4\sqrt5\,\pi T_\star} \frac{\partial T_{21}}{\partial\delta} \sigma_\delta.
\end{equation}
At e.g. $z=40$, typical values of $\partial T_{21}/\partial\delta$ and $\sigma_\delta$ are $-1$ K (remember the factor of $1+z$ since we want the temperature perturbations at $z=40$) and 0.1, respectively \cite{2004PhRvL..92u1301L}; this leads to a root-mean-square frequency splitting of $1.5\times 10^{-16}$ s$^{-1}$. This is two orders of magnitude smaller than the splitting coming from the CMB anisotropy, and hence is neglected here. We also note that the 21 cm self-induced quadrupole should be further distinguishable from a gravitational wave signal, since it peaks on smaller scales (it ``inherits'' the shape of the density power spectrum, with no factors of $k$). Moreover, it is locally sourced, and does not have a preferred direction coherent over large scales in the same sense as the CMB quadrupole-induced polarization.

\subsection{Kinematic quadrupole}
\label{ss:kinematic}

The motion of baryonic gas relative to the CMB rest frame leads to a dipole intensity perturbation at linear order as measured in the baryon rest frame. However, a dipole intensity perturbation does not split the $M_F$ sub levels of hydrogen -- only a quadrupole perturbation does that. At second order in the baryon-radiation relative velocity, however, the baryons see a ``kinematic quadrupole'' due to second-order terms in the Doppler shift formula \cite{1980MNRAS.190..413S}; see Ref.~\cite{2000ApJ...529...12H} for an extensive discussion in the context of secondary CMB anisotropies.

The splitting due to the kinematic CMB quadrupole can easily be computed. Let us first consider the case of a gas parcel moving at velocity $\beta c$ in the $z$-direction. The squared CMB temperature seen by that parcel in direction $\hat{\bm n}$ is
\begin{equation}
\left( \frac{T({\bm n})}{T_\gamma} \right)^2
= 1 + 2\beta P_1(\hat n_3) + 2\beta^2 P_2(\hat n_3) + {\cal O}(\beta^3).
\end{equation}
Recall that the dynamic magnetic dipole splitting of the $M_F$ levels is proportional to radiation temperature squared, so we should consider the quadrupole moment of $T^2$ rather than some other power. This is equivalent to a an anisotropy $a_{20}^{\rm kin} = \sqrt{4\pi/5}\,\beta^2$. Generalizing to arbitrary $\beta$ gives
\begin{equation}
a_{2m}^{\rm kin} = \sqrt{6\pi} \sum_{m'm''} (-1)^m \threej211{-m}{m'}{m''} \beta_{m'} \beta_{m''},
\label{eq:kin}
\end{equation}
where the polar components of $\beta_m$ have been used, the form with the $3j$ symbol is required by spherical symmetry, and the pre factor was chosen to re-produce the specific example considered above. The power spectrum corresponding to the kinematic quadrupole is
\begin{equation}
C_2^{\rm kin} = \langle |a_{2m}^{\rm kin}|^2 \rangle = \frac{4\pi}{15} \beta_{\rm rms}^4,
\label{eq:kqC2}
\end{equation}
where $\beta_{\rm rms}$ is the root-mean-square baryon velocity relative to the CMB (summed over all axes: i.e. $\langle\beta_{m'}^\ast\beta_{m''}\rangle = \beta_{\rm rms}^2\delta_{m'm''}/3$) and we have assumed a Gaussian velocity distribution so that Wick's theorem applies in the simplification of Eq.~(\ref{eq:kqC2}).

The root-mean-square velocity is given in linear perturbation theory by
\begin{equation}
\beta_{\rm rms} = \sqrt{\int \frac{dk}k\,\Delta^2_\delta(k)\,\left( \frac{faH}{k} \right)^2}
\end{equation}
and should scale in linear perturbation theory as $\propto (1+z)^{-1/2}$. Using the Fisher matrix code of Ref.~\cite{2009arXiv0901.0721A}, the above integral evaluates to $5.3\times 10^{-4}$ at $z=19$. We thus conclude from Eq.~(\ref{eq:kqC2}) that
\begin{equation}
C_2^{\rm kin} = 6.4\times 10^{-14}\left( \frac{1+z}{20} \right)^{-2}.
\end{equation}
As compared to the primordial quadrupole of $C_2 = \Delta_\zeta^2/25 = 10^{-10}$, this is smaller by a factor of 1600 (in power).

However, the expected contamination to the primordial gravitational wave signal is much lower since the velocity perturbations are dominated by small scales -- of order $k\sim k_{\rm eq}$ -- whereas the CMB perturbations (including those from tensor modes) are dominated by the horizon scale, $k\sim aH$. To determine the kinematic quadrupole fluctuations on large scales, we write the power spectrum,
\begin{equation}
\langle a_{2m}^{{\rm kin}\ast}({\bf k})a_{2m'}^{{\rm kin}\ast}({\bf k}) \rangle = (2\pi)^3 \delta_{mm'}
P_{\rm kin}^{(m)}(k) \delta^{(3)}({\bf k}-{\bf k}'),
\end{equation}
where we set ${\bf k}$ to be on the $z$-axis. The power spectra $P_{\rm kin}^{(m)}(k) = P_{\rm kin}^{(-m)}(k)$ by parity. Since $a_{2m}^{{\rm kin}}$ is a simple product of velocities in real space, its power spectrum is an auto-convolution of the velocity power spectrum, which is the density power spectrum multiplied by factors of $faH/k_1$ and with a factor of $\hat{\bf k}_1$:
\begin{eqnarray}
P_{\rm kin}^{(m)}(k) &=& 12\pi
\int \frac{d^3{\bf k}_1}{(2\pi)^3}
\left( \frac{faH}{k_1} \right)^2
\left( \frac{faH}{k_2} \right)^2
\nonumber \\ && \times
P_\delta(k_1) P_\delta(k_2)
\sum_{m_1m_2m_3m_4}
\threej211{-m}{m_1}{m_2}
\nonumber \\ && \times
\threej211{-m}{m_3}{m_4}
[\hat{\bf k}_1]_{m_1}^\ast
[\hat{\bf k}_2]_{m_2}^\ast
[\hat{\bf k}_1]_{m_3}
[\hat{\bf k}_2]_{m_4},
\nonumber \\ &&
\end{eqnarray}
where ${\bf k}_2\equiv {\bf k}-{\bf k}_1$. [The factor of $12\pi$ comes from the $\sqrt{6\pi}$ in Eq.~(\ref{eq:kin}) and a combinatorial factor of 2.] In the limit of $k\ll k_{\rm eq}$, we may approximate ${\bf k}_2\approx -{\bf k}_1$ and this simplifies to
\begin{eqnarray}
P_{\rm kin}^{(m)}(k) &=& 12\pi
\int \frac{d^3{\bf k}_1}{(2\pi)^3}
\left( \frac{faH}{k_1} \right)^4
[P_\delta(k_1)]^2
\nonumber \\ && \times
\left| \sum_{m_3m_4}
\threej211{-m}{m_3}{m_4}
[\hat{\bf k}_1]_{m_3}
[\hat{\bf k}_1]_{m_4} \right|^2,
\nonumber \\ &=& \frac{4}{25\pi}
\int k_1^2
\left( \frac{faH}{k_1} \right)^4
[P_\delta(k_1)]^2\,dk_1,
\label{eq:744}
\end{eqnarray}
where in the second equality we performed the angular integral over $\hat{\bf k}_1$, leaving only the radial integral explicit. (The angular average of the square norm in the first expression is $2/75$.) This integral is independent of $m$ and evaluates to $\lim_{k\rightarrow 0}P_{\rm kin}^{(m)}(k) = 1.9\times 10^{-7}\,$Mpc$^3$ at $z=19$.

For a white noise spectrum (independent of $k$), and taking into account the 5 possible values of $m=-2, ...+2$, the total variance coming from scales $<k_{\rm cut}$ is $(5k_{\rm cut}^3/6\pi^2)P(k)$. This leaves
\begin{equation}
C_2^{\rm kin}(<k_{\rm cut}) = 6\times 10^{-17} \left( \frac{k_{\rm cut}}{2.8aH} \right)^3
\end{equation}
at $z=19$. We have scaled $k_{\rm cut}$ relative to the wavenumber $2.8aH$, since modes with $k<2.8aH$ contribute 90\%\ of the variance of the CMB quadrupole. This suggests that at the horizon scale, $k_{\rm cut}/(aH)\sim {\cal O}(1)$, the kinematic quadrupole is 6 orders of magnitude below the CMB quadrupole, and hence would only become important for an experiment capable of probing tensor-to-scalar ratios of ${\cal O}(10^{-6})$. This leaves aside the fact that the kinematic quadrupole is derived from the scalar perturbations and hence it should be possible to predict it from the density field measured in 21 cm experiments.

\subsection{Starlight}
\label{ss:starlight}

At the lower redshifts, which are also the most observationally accessible, the spin temperature of the hydrogen atoms is likely to be ``pumped'' to $T_s\approx T_k$ by Lyman-$\alpha$ radiation \cite{1952AJ.....57R..31W, 1958PIRE...46..240F, 1959ApJ...129..536F, 2006MNRAS.367..259H, 2006MNRAS.367.1057P}. However, during this epoch the ambient radiation field of the starlight -- which is probably much more anisotropic than the CMB -- will lead to a radiatively-induced splitting of the hydrogen $F=1$ level. In order to assess the suitability of this epoch for studies of primordial gravitational waves, we need to determine the order of magnitude of this effect.

We first note that, according to our calculations of the energy splittings induced by anisotropic blackbodies, that the electric dipole splitting will dominate at temperatures exceeding $\sim 600$ K or photon wavelengths $\lambda\lesssim 5$ $\mu$m, because of the additional two powers of frequency. (This ultimately results from the fact that the magnetic dipole operator connects the $1s$ levels to each other, whereas the electric dipole operator only connects them to electronically excited states $np$.) Thus for our studies of starlight-induced splitting, we focus on the electric dipole rather than the magnetic dipole splitting. This splitting is proportional to the covariance matrix of the electric field, and hence to the total energy density multiplied by the anisotropy. If we assume an order-unity anisotropy, then we should estimate the order of magnitude of the electric dipole splitting by the replacement in Eq.~(\ref{eq:ans-ed}):
\begin{equation}
a_{\rm rad}T_\gamma^4 a_{20} \rightarrow n_{\rm b} \epsilon_\star,
\end{equation}
where $\epsilon_\star$ is the energy in starlight per baryon.
This leads to
\begin{equation}
\frac{\Delta{\cal E}^{\rm e.d.}_{11}-\Delta{\cal E}^{\rm e.d.}_{10}}{\hbar}
\sim 10^{-19}\,{\rm s}^{-1}\,\left( \frac{1+z}{20}\right)^3 \frac{\epsilon_\star}{\rm eV},
\end{equation}
where the $1+z$ scaling arises since we normalized the starlight energy to the number of baryons. The evolution of $\epsilon_\star$ is uncertain, but it is estimated that the Lyman-$\alpha$ coupling becomes saturated (in the sense of $\tilde x_\alpha\sim 1$) when there are 3 eV of starlight per baryon per $\ln\nu$ in the vicinity of Lyman-$\alpha$ \cite{2004ApJ...602....1C}. If a fraction of order unity of the starlight from early galaxies emerges in the far-ultraviolet, then, it is reasonable to expect an energy splitting of $\lesssim 10^{-18}$ s$^{-1}$ due to starlight when $\tilde x_\alpha\sim 1$ (we use the $\lesssim$ sign here since the anisotropy of the radiation may be less than of order unity). This is $\sim 4$ orders of magnitude less than the expected energy splitting from the CMB, and hence we neglect it.

\section{Depolarization due to Lyman-$\alpha$ scattering}
\label{sec:lya}

\changetext{In this section, we consider neutral hydrogen atoms (in their ground $1s$ electronic state) immersed inside an isotropic and unpolarized Lyman-$\alpha$ radiation field. The resonant scattering of the photons within the Lyman-$\alpha$ line causes two-step transitions between the hyperfine sublevels. We will write down the resulting evolution of the atomic density matrix within the hyperfine basis, and infer the rate of depolarization of aligned states (i.e., states with nonzero $\mathscr{P}_{1m}$).}

\changetext{Note that due to the above assumptions, there is no tensor of spin greater than zero that we can form from the incident Lyman-$\alpha$ radiation field. If we neglect stimulated emission from the short-lived intermediate state in the scattering process, the rates do not depend on the outgoing radiation. Hence, Lyman-$\alpha$ scattering can only connect the spherical components of the atomic density matrix (the $\mathscr{P}_{jm}$ of Eq.~\eqref{eq:pjm}) with the same $j$ and $m$.}

\changetext{In the space of the hyperfine sublevels of the $1s$ state, the perturbation due to resonant scattering is similar to the Hamiltonian of Eq.~\eqref{eq:aC1}. The relevant terms are those with the intermediate state (labeled by the index $n$ in Eq.~\eqref{eq:aC1}) within the $2p$ state. In our application, we replace the energy $\mathcal{E}_n \rightarrow \mathcal{E}_n - i \Gamma_n$, where the width $\Gamma_n$ accounts for the finite lifetime of the $2p$ states.}

\changetext{The interaction matrix element between the atom and a photon is
\begin{equation}
  \langle a \vert H_{\rm int} \vert i \rangle  = -{\mathbf d}_{ai} \cdot {\mathbf E}  = - i \sqrt{2\pi\omega} \, {\mathbf e} \cdot {\mathbf d}_{ai}, \label{eq:matrixelements}
\end{equation}
where $\mathbf e$ and $\mathbf E$ are the polarization and electric field of the photon, and $\mathbf d$ is the transition dipole moment. In this equation (and in the rest of the section), we omit all factors of $\hbar$ and $c$.} 

\changetext{The cross-section for the transition between levels $i \rightarrow f$ is given by the Fermi Golden rule:
\begin{align}
  d\sigma_{i \rightarrow f} & = 2\pi \vert \Delta \mathcal{E}_{f i} \vert^2 \frac{\omega_B^2\, d^2\hat{\mathbf n}_B}{(2\pi)^3}, \label{eq:fermig}
\end{align}
where $\omega_B$ and $\hat{\mathbf n}_B$ are the energy and direction of the outgoing photon, the final term is its density of states, and the term $\Delta \mathcal{E}_{f i}$ is obtained by substituting Eq.~\eqref{eq:matrixelements} into Eq.~\eqref{eq:aC1}.\footnote{\changetext{This derivation is standard, and we have omitted the intermediate steps. The reader might be confused by the fact that both the perturbation of Eq.~\eqref{eq:aC3}, and the rate for a process with a cross-section (which is nominally the square of the perturbation) are both linear in the flux of the incident radiation. The resolution is that in the derivation of the transition rate, the average over the radiation field should be performed {\em after} squaring the matrix elements (for more information, see, e.g., \S61 of Ref.~\cite{Lifshitz71}, or the derivation of Eq. (III,11) of Ref.~\cite{Barrat61}).}}}

\changetext{For the purposes of fixing the notation, we first write down the net cross-section in the case where the initial atom is not polarized, i.e., only the net population $\mathscr{P}_{00}$ is nonzero. The initial state can have either $F_i = 0$ or $1$. Expanding out the matrix element in Eq.~\eqref{eq:fermig}, we get
\begin{align}
  d\sigma_{i \rightarrow f} & = \omega_A \omega^3_B c^{\mu \nu \alpha \beta}_{i \rightarrow f} \bar{e}^B_\mu {e}_\nu^A e_\alpha^B \bar{e}_\beta^A\, d^2\hat{\mathbf n}_B,
 \label{eq:scattavg}
\end{align}
where
\begin{align}
  c^{\mu \nu \alpha \beta}_{i \rightarrow f} & = \frac{1}{2F_i + 1} e^4 \sum_{m_i, m_f} \sum_{a, b} \nonumber \\
  & ~~~ \frac{\langle f | r^\mu | a \rangle \langle a | r^\nu | i \rangle \langle i | r^\beta | b \rangle \langle b | r^\alpha | f \rangle}{\left(\omega + \omega_i - \omega_a + i \Gamma_a/2\right)\left(\omega^\prime + \omega_f - \omega_b - i \Gamma_b/2\right)}.
  \label{eq:scattavgtensor}
\end{align}
In the above equation, the subscript $_{A}$ refers to the incoming photon, the unsubscripted letter $e$ is the charge of the electron (not to be confused with the polarization of the photon, whose components are always subscripted), and Greek indices indicate the spherical coordinate system, conjugates in which are indicated by bars. The sums over the azimuthal quantum numbers are equivalent to averaging over the orientation of the spin of the initial atom, and summing over that of the final atom.}

\changetext{The average over the polarization of the incoming photon, and the sum over the polarization and direction of the outgoing photon lead to the replacement
\begin{align}
  \bar{e}^B_\mu {e}_\nu^A e_\alpha^B \bar{e}_\beta^A \,d^2\hat{\mathbf n}_B & \rightarrow \frac{8\pi}{3} g_{\mu \alpha} g_{\nu \beta}, \label{eq:polavg}
\end{align}
where $g_{\mu \nu} = (-1)^\mu \delta_{\mu, -\nu}$ is the metric tensor in the spherical coordinate system.} 

\changetext{Next, we need to generalize Eq.~\eqref{eq:scattavg} to the case where the initial and final states of the atoms are polarized, i.e., some $\mathscr{P}_{jm}$ with $j \neq 0$ is nonzero. An inspection of Eq.~\eqref{eq:scattavgtensor} points the way forward: we need to replace the uniform sums over the azimuthal quantum numbers $m_i$ and $m_j$ with weighted sums, with coefficients that project out the spherical components $\mathscr{P}_{jm}$ from the density matrix components $\rho_{m_1 m_2}$.}

\changetext{We can read off these coefficients from the definition in Eq.~\eqref{eq:pjm} and its inverse. We write these down for a general level with total angular momentum $F$:}
\begin{widetext}
\changetext{
\begin{equation}
  \mathscr{P}_{jm}^{F} = \sqrt{(2j+1)(2F+1)} \sum_{m_1, m_2} (-1)^{F-m_2} \tj{F}{j}{F}{-m_2}{m}{m_1} \rho_{F m_1, F m_2} \label{eq:pjmf}
\end{equation}
and
\begin{equation}
  \rho_{F m_1, F m_2} = \sum_{j m}\sqrt{\frac{2j+1}{2F + 1}} (-1)^{F-m_2} \tj{F}{j}{F}{-m_2}{m}{m_1} \mathscr{P}_{j m}. \label{eq:pjminvf}
\end{equation}
One final identity we need is the decomposition of the combination of metric tensors in Eq.~\eqref{eq:polavg} into spin-$K$ irreducible parts, which facilitates the rest of the calculation (see Ref.~\cite{2006MNRAS.367..259H}):
\begin{align}
\label{eq:spink}
g_{\mu \alpha} g_{\nu \beta} = \sum_{K=0}^2 \Pi^{(K)}_{\mu \nu \alpha \beta},~~{\rm where}~~
  \Pi^{(K)}_{\mu \nu \alpha \beta} 
    = (2K+1) \sum_{m_K} (-1)^{K-m_K} \tj{1}{1}{K}{\mu}{\nu}{m_K} \tj{1}{1}{K}{\alpha}{\beta}{-m_K}.
\end{align}
Now we have all the pieces needed to calculate the cross section $\sigma_{i \rightarrow f, (j)}$ for the scattering of the $j^{\rm th}$ spherical moment moment from the level $i$ to $f$ (i.e., $\mathscr{P}_{jm}^{F_i} \rightarrow \mathscr{P}_{jm}^{F_j}$). First, we begin with Eq.~\eqref{eq:scattavgtensor}, in which we replace the sums over $m_i$ and $m_j$ with weights chosen from Eqs.~\eqref{eq:pjmf} and \eqref{eq:pjminvf} to project out the $(j,m)$ moment in both the initial state $i$, and final state $j$. Second, we use the Wigner-Eckart theorem for all the matrix elements in Eq.~\eqref{eq:scattavgtensor}. Third, we average and sum over the polarizations of the initial and final photons, respectively using Eq.~\eqref{eq:polavg}. Finally, we substitute Eq.~\eqref{eq:spink} for the metric tensors, and write the required cross section as
\begin{align}
  \sigma_{i \rightarrow f,(j)} & = \frac{8\pi}{9} \omega_A \omega_B^3 \sum_K \bar{G}_{i \rightarrow f,(j)}^{(K)}. \label{eq:dmevolj}
\end{align}
The right hand side is a sum over irreducible spherical components (of spin-K) of this cross-section, each of which is
\begin{align}
  \bar{G}_{i \rightarrow f,(j)}^{(K)}
  & = \sqrt{2j+1} \sqrt{\frac{2F_{f}+1}{2F_{i}+1}} (2K+1) e^{4} \sum_{\mu, \nu, \alpha, \beta, M_K, m_1, m_2, m_1^\prime, m_2^\prime, a, b, m_a, m_b, j^\prime, m^\prime} \!\!\!\!\!\!\!\! \sqrt{2j^\prime + 1} 
  \frac{\langle f \| \bm{r} \| a \rangle \langle a | \bm{r} \| i \rangle \langle i \| \bm{r} \| b \rangle \langle b \| \bm{r} \| f \rangle}{\left(\Delta\omega_{ai} + i \Gamma_a/2\right)\left(\Delta\omega_{bi} - i \Gamma_b/2\right)}  \notag \\
  & \hspace{5pt}\times (-1)^{K-M_K} \tj{1}{1}{K}{\mu}{\nu}{M_K} \tj{1}{1}{K}{\alpha}{\beta}{-M_K}
  (-1)^{F_f-m_2} \tj{F_f}{j}{F_f}{-m_2}{m}{m_1} (-1)^{F_i-m_2^\prime} \tj{F_i}{j^\prime}{F_i}{-m_2^\prime}{m^\prime}{m_1^\prime}  \notag \\
  & \hspace{5pt} \times (-1)^{F_f - m_1} \tj{F_f}{1}{F_a}{-m_1}{\mu}{m_a} (-1)^{F_a - m_a} \tj{F_a}{1}{F_i}{-m_a}{\nu}{m_1^\prime} 
   \notag \\
  & \hspace{5pt} \times 
  (-1)^{F_i - m_2^\prime} \tj{F_i}{1}{F_b}{-m_2^\prime}{\beta}{m_b} (-1)^{F_b - m_b} \tj{F_b}{1}{F_f}{-m_b}{\alpha}{m_2}.
\end{align}
The double-barred symbols are the reduced matrix elements of the position operator. This expression can be simplified using the definition of the $6j$ symbol:
\begin{align}
\bar{G}_{i \rightarrow f,(j)}^{(K)} 
  & = \sqrt{\frac{2F_{f}+1}{2F_{i}+1}} (2K+1) e^{4} \sum_{a, b}  \frac{\langle f \| \bm{r} \| a \rangle \langle a | \bm{r} \| i \rangle \langle i \| \bm{r} \| b \rangle \langle b \| \bm{r} \| f \rangle}{\left(\Delta\omega_{ai} + i \Gamma_a/2i\right)\left(\Delta\omega_{bi} - i \Gamma_b/2\right)}  (-1)^{K+j}
  \notag \\ & \hspace{5pt} \times \sj{K}{F_i}{F_f}{F_a}{1}{1} \sj{K}{F_f}{F_i}{F_b}{1}{1} \sj{j}{F_i}{F_i}{K}{F_f}{F_f}.
\end{align}
Given this result, we can perform the sum over the index $K$ in Eq.~\eqref{eq:dmevolj} and use the symmetries of the reduced matrix elements to obtain
\begin{equation}
\sum_K \bar{G}_{i \rightarrow f,(j)}^{(K)} 
   = \sqrt{\frac{2F_{f}+1}{2F_{i}+1}} e^{4} \sum_{a, b} \frac{\langle a \| \bm{r} \| f \rangle^* \langle a \| \bm{r} \| i \rangle \langle b \| \bm{r} \| i \rangle^* \langle b \| \bm{r} \| f \rangle}{\left(\Delta\omega_{ai} + i \Gamma_a/2\right)\left(\Delta\omega_{bi} - i \Gamma_b/2\right)}  (-1)^{F_i - F_f} \sj{F_a}{F_b}{j}{F_i}{F_i}{1} \sj{F_a}{F_b}{j}{F_f}{F_f}{1}. \label{eq:sumK}
\end{equation}
The Lyman-$\alpha$ line is split into six lines due to spin-orbit and hyperfine corrections; their parameters are given in Table B1 of Ref.~\cite{2006MNRAS.367..259H}. Exactly following their treatment, we can express the reduced matrix elements in terms of the (identical) HWHM of these lines $\gamma_{2p} = \Gamma_{2p}/4\pi$, and the terms involving the frequency offsets lead to the line and interference profiles, $\phi_{AA}(\nu)$ and $\phi_{AB}(\nu)$.}

\changetext{We substitute the result of Eq.~\eqref{eq:sumK} into Eq.~\eqref{eq:dmevolj}, and set $\omega_A \approx \omega_B$ in the prefactor. If we take $j=0$, this gives the cross-sections in the unpolarized case, which were computed in Ref.~\cite{2006MNRAS.367..259H}. We list here the relevant unpolarized cross-sections:
\begin{align}
  \sigma_{1 \rightarrow 1} & =  \frac{3}{2} \lambda_{\text{Ly}\alpha}^2 \gamma_{2p} \left( \frac{1}{9}\phi_{AA} + \frac{4}{27}\phi_{BB} + \frac{1}{27}\phi_{DD} +  \frac{5}{9}\phi_{EE} + \frac{4}{27}\phi_{BD} \right) \, {\rm and} \notag \\
  \sigma_{1 \rightarrow 0} & = \frac{3}{2} \lambda_{\text{Ly}\alpha}^2 \gamma_{2p} \left(\frac{2}{27}\phi_{BB} + \frac{2}{27}\phi_{DD} - \frac{4}{27}\phi_{BD} \right).
\end{align}
Both the cross-sections and line profiles are functions of frequency, which we have suppressed in the above equations. In the language of V17, these cross-sections together give the depopulation rate of the $F=1$ level.}

\changetext{The main new result in this section is the cross-section for scattering the $j=1$ moment from the $F=1$ level to itself (or the repopulation rate, in the language of V17):
\begin{equation}
  \sigma_{1 \rightarrow 1, (1)} 
  = \frac{3}{2} \lambda_{ {\rm Ly}\alpha}^2 \gamma_{2p} \left( \frac{1}{27} \phi_{BB} + \frac{1}{108} \phi_{DD} + \frac{5}{12} \phi_{EE} + \frac{4}{27} \phi_{AB} + \frac{2}{27} \phi_{AD} + \frac{1}{27} \phi_{BD} + \frac{5}{27} \phi_{BE} + \frac{5}{54} \phi_{DE} \right).
\end{equation}
Putting everything together, the net depletion rate of the orientation $\mathscr{P}_{1m}$ due to Lyman-$\alpha$ scattering is
\begin{align}
\biggl. \frac{d \mathscr{P}_{1 m}}{d t} \biggr\rvert_{{\rm Ly}\alpha}  & = - 4\pi \int d\nu \, J(\nu) \left( \sigma_{1 \rightarrow 1} + \sigma_{1 \rightarrow 0} - \sigma_{1 \rightarrow 1, (1)} \right) \mathscr{P}_{1 m} \notag \\
  & = - 6\pi \lambda_{ {\rm Ly}\alpha}^2 \gamma_{2p} J(\nu_{{\rm Ly}\alpha}) \mathscr{P}_{1 m} \int d\nu \,  \frac {J(\nu)}{J(\nu_{{\rm Ly}\alpha})} \notag \\
  & ~~~\times \left( \frac{1}{9}\phi_{AA} + \frac{5}{27} \phi_{BB} + \frac{11}{108} \phi_{DD} + \frac{5}{36} \phi_{EE} - \frac{4}{27} \phi_{AB} - \frac{2}{27} \phi_{AD} - \frac{1}{27} \phi_{BD} - \frac{5}{27} \phi_{BE} - \frac{5}{54} \phi_{DE} \right) .
\end{align}
If we neglect the spectral distortion in the core of the line, the flux ratio inside the integrand equals unity. Under this assumption, the remainder of the integrand is composed solely of known line profiles; we use the line parameters from Ref.~\cite{2006MNRAS.367..259H} to numerically compute the integral, which evaluates to $0.445$. This is the source of the extra prefactor in Eq.~\eqref{eq:lya}.}
\end{widetext}

\bibliography{forward,forward_teja}

\end{document}